\documentclass[twocolumn,tighten]{aastex7}
\shorttitle{SpiderCat}
\shortauthors{K. I. I. Koljonen \& M. Linares}
\begin{document}

\title{SpiderCat: A Catalog of Compact Binary Millisecond Pulsars}

\author[orcid=0000-0002-9677-1533]{Karri I. I. Koljonen}
\affiliation{Institutt for Fysikk, Norwegian University of Science and Technology, H\"{o}gskoleringen 5, Trondheim, 7491  Norway}
\email[show]{karri.koljonen@ntnu.no}  

\author[orcid=0000-0002-0237-1636]{Manuel Linares} 
\affiliation{Institutt for Fysikk, Norwegian University of Science and Technology, H\"{o}gskoleringen 5, Trondheim, 7491  Norway}
\affiliation{Departament de F{\'i}sica, EEBE, Universitat Polit{\`e}cnica de Catalunya, Av. Eduard Maristany 16, E-08019 Barcelona, Spain}

\email[show]{manuel.linares@ntnu.no}

\begin{abstract}

We present SpiderCat, a multiwavelength catalog of all publicly known compact binary millisecond pulsars (MSPs) in the Galactic field. These systems, colloquially known as ``spiders,'' consist of neutron stars in tight orbits with low-mass companions, which are gradually ablated by the pulsar wind. SpiderCat includes both primary subclasses---redbacks and black widows---distinguished by companion mass, as well as candidates and peculiar systems such as transitional, huntsman, and tidarren MSPs. As of this initial release, SpiderCat contains 111 entries: 30 redbacks, 50 black widows, two huntsmans, 23 redback candidates, five black widow candidates, and one huntsman candidate. In this paper, we compile and summarize key parameters for each system, including spin and orbital properties, and multiwavelength data from radio, optical, X-ray, and $\gamma$-ray observations. An interactive, publicly accessible web interface, at \url{https://astro.phys.ntnu.no/SpiderCAT}, enables exploration and visualization of the data. The rapid growth of the number of known spiders, accelerated by the Fermi Large Area Telescope survey and its ability to identify MSPs in $\gamma$-rays, has opened the door to population-level studies. Utilizing SpiderCat, we analyze trends in spin period, orbital period, companion mass, emission properties, and spatial distribution. SpiderCat serves as a dynamic, multiwavelength repository for this unique class of binary pulsars, facilitating new discoveries and constraints on pulsar evolution, particle acceleration, and the neutron star equation of state.
\end{abstract}


\section{Introduction} 

%ML struct:
%
%LAT, MSPs
%MSPs were thought to be weak gamma-ray emitters before Fermi-LAT was launched
Before the Fermi $\gamma$-ray Space Telescope was launched in 2008, only seven $\gamma$-ray pulsars had been confidently detected (by the Energetic $\gamma$-Ray Experiment Telescope (EGRET), between 1991 and 2000), all of which were young and slowly rotating \citep[spin period $P>30$~ms;][]{Thompson08}.
%3PC:
The Large Area Telescope (LAT) onboard Fermi, with an effective area nearly 10 times larger than EGRET's, has proven to be an efficient pulsar-discovery machine \citep{FermiLAT09}.
Based on 12 yr of all-sky survey LAT data, Fermi's Third Pulsar Catalog (3PC) includes 294 $\gamma$-ray pulsars \citep[and 45 candidates;][]{saa+23}.
%MSPs
Roughly half of these 3PC $\gamma$-ray pulsars are millisecond pulsars (MSPs): ``old," rapidly spinning ($P<30$~ms) neutron stars in binary systems, with relatively low magnetic fields ($B\sim10^8-10^9$~G), that have been ``recycled'' through the accretion of matter \citep{Alpar82}.
%NEED NEXT?
Because MSPs have a higher Galactic scale height than young pulsars \citep{lorimer13}, they are more easily discovered in LAT surveys far from the Galactic plane, where $\gamma$-ray diffuse emission does not obscure faint point sources.

%SPIDER BOOM
Many of the new Fermi-LAT pulsar discoveries have been identified as compact binary MSPs, characterized by short orbital periods (typically $P_{\rm b}\lesssim1$~d) and nondegenerate or semidegenerate companion stars (see definitions in Section~\ref{sec:definition}).
In such compact systems, interactions between the pulsar and its companion can be extreme; the outer layers of the companion star may be ablated leading to occultations or ``eclipses" of the radio pulsar \citep{fst88}.
These destructive effects have inspired the cannibalistic spider nicknames for compact binary MSPs---such as black widows (BWs) and redbacks (RBs)---commonly referred to in the literature (and hereinafter) as spider pulsars, spider MSPs, or simply ``spiders" \citep[e.g.,][]{Roberts11}.

Pulsars represent the most numerous Galactic class of $\gamma$-ray sources.
They are characterized by low $\gamma$-ray variability and high spectral curvature, which sets them apart from blazars (the most numerous extragalactic class) that present typically higher variability and power law spectra. 
These characteristics enable an efficient selection of pulsar candidates among thousands of unidentified Fermi-LAT sources \citep[$>$7000 cataloged sources, with $>$2000 still unidentified;][]{aab+22}.
Thus, the Fermi-LAT point source catalogs have been a treasure trove for spider discoveries over the past 15 yr (Section~\ref{sec:discoveries}).

%tMSPs, states
An ``elite'' subset of compact binary MSPs (belonging to the RB subtype, see below) have been observed in the so-called disk state, where they form an accretion disk and the radio pulsar becomes undetectable \citep{Wang09,deMartino15}.
These so-called transitional MSPs (tMSPs; labeled ``tr'' below) show optical disk emission lines and a characteristic X-ray variability pattern known as X-ray mode switching \citep{Linares14}.
In the pulsar state, when the radio pulsar is detected, spiders show lower luminosities and orbital modulation (with different origins) throughout the electromagnetic spectrum, from radio to $\gamma$-rays.
These properties are used to identify new spiders, as described in Section~\ref{sec:discoveries}.

%intrabinary shock, particle acceleration
As a consequence of the compact orbits, an intrabinary shock (IBS) is thought to form between the pulsar and companion winds in spider MSPs \citep{arons93,phinney88}.
This shock produces non-thermal (synchrotron) X-ray emission that is modulated with the orbit \citep[e.g.,][]{stappers03,huang12,romani16,wadiasingh17,vandermerwe20}.
The IBS can also be an efficient site for particle acceleration \citep[e.g.,][]{Harding90,sironi11,cortes22,cortes24,cortes25}, potentially contributing to the cosmic-ray positrons detected on Earth \citep{linares21}.
%
%IBS, particles
%neutron star masses
During the gigayear-long recycling process, neutron stars in low-mass X-ray binaries can accrete up to about 1~M$_\odot$ \citep[e.g.,][]{misra25a}.
As a result, MSPs in general---and spiders in particular---are thought to harbor massive neutron stars.
Indeed, several studies have found evidence for massive neutron stars from dynamical measurements of compact binary MSPs \citep[e.g.,][]{Linares18b,Strader19,Romani22}.

For the reasons above, the field of compact binary MSPs has experienced a boom in the last decade.
%
%%% WHY DO WE NEED A CATALOGUE:
We now know of more than 100 spiders, enabling meaningful population studies.
While existing pulsar catalogs (Section~\ref{sec:sample}) focus on radio timing properties, there is no public, up-to-date repository of multiwavelength information for spider MSPs (to the best of our knowledge).
Thus, the time is ripe to collect and analyze the spin, orbit and multiwavelength properties of all known Galactic-field spiders.
This is the goal of the present work.

%END ML STRUCT

\subsection{Spider Discoveries}
\label{sec:discoveries}

\begin{figure*}
\plotone{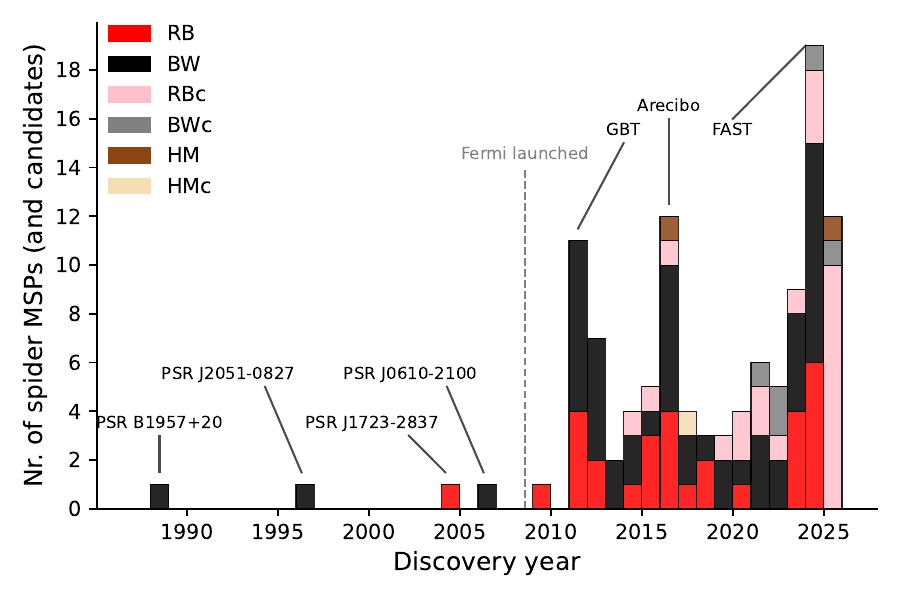}
\caption{Histogram showing the discovery year of spider systems included in SpiderCat, color coded by spider subclass. The discovery year is defined as the year in which a paper was published confirming the system as a spider or spider candidate, or reporting the detection of millisecond pulsations from the source. The first four spider systems are highlighted. The dramatic increase in discoveries following the launch of the Fermi $\gamma$-ray Space Telescope reflects its critical role in guiding targeted radio (e.g., 350~MHz Green Bank Telescope (GBT) and 327~MHz Arecibo surveys) and multiwavelength surveys.
\label{fig:disc_year}}
\end{figure*}

Thanks to a flurry of Fermi-driven discoveries, the spider population has grown from a rara avis (with only four systems known before 2008; see Figure~\ref{fig:disc_year}) into an important subclass of MSPs (109 systems presented here as of 2025).
Indeed, spiders now constitute 15\%--20\% of the 552 MSPs currently known in the Galactic disk \citep[][where the higher end of the range includes candidate systems]{GalMSPs}.
The methods that have been used to discover new spiders include the following.

\begin{enumerate}

    \item {\it Wide-field radio timing surveys (e.g., the Galactic Plane Pulsar Snapshot (GPPS) survey with the Five-hundred-meter Aperture Spherical Telescope (FAST); \citealt{why+24}; see also \citealt{fsk+04})}. These surveys typically observe each sky location once or only a few times, with relatively short integrations. Because in spider systems the pulsar is occulted during a large fraction of the orbit (20\%--60\%), these surveys are suboptimal for discovering them. Follow-up observations are required to measure orbital parameters. 
    
    \item {\it $\gamma$-ray-guided radio timing searches (e.g., by the Pulsar Search Consortium using the GBT; \citealt{hrm+11,rap+12}; see also \citealt{bgc+13})}. This method targets specific unidentified $\gamma$-ray sources using longer and/or more numerous radio pointings, after selecting pulsar candidates based on their $\gamma$-ray properties (as discussed above). These searches have been particularly effective in discovering new spiders, which are readily identified by their radio ``eclipses'' or occultations.

    \item {\it $\gamma$-ray-guided optical searches (e.g., the Compact Binary Pulsar Search; \citealt{tlk+24}; see also \citealt{rs11,Kong12,Linares17,lrl+25})}. Compared to the more common binary MSPs with white dwarf (WD) companions, spider MSPs with main-sequence and brown dwarf companions are typically more luminous in the optical--IR band. Their compact orbits result in large-amplitude orbital modulations due to the gravitational distortion and irradiation of the companion star, as viewed at different orbital phases. These features enable optical--IR searches for spider candidates, especially the optically brighter RB subclass (Sections~\ref{sec:definition} and \ref{sec:sample}), by targeting their characteristic flux (and radial velocity) variations. The candidate sources are typically selected based on $\gamma$-ray properties (low variability and highly curved spectra).

    \item {\it Blind $\gamma$-ray searches (e.g., \citealt{pgf+12}; see also \citealt{nck+20})}. The Fermi-LAT has been surveying the full sky since 2008, providing a rich dataset for blind pulsar searches. These searches fold $\gamma$-ray photons over a range of spin and orbital parameters to detect pulsations from a given source or sky location. Although computationally intensive, especially for MSPs with unknown orbital parameters, they can be aided by prior knowledge of such parameters, e.g., the orbital period \citep{cnv+21}. In addition, some spiders have shown orbital modulation \citep{An20} and eclipses \citep{Clark23} in their $\gamma$-ray light curves.

    \item {\it X-ray searches (e.g., \citealt{bh15}; see also \citealt{cpd+19})}. X-ray properties can also help identify new spider systems. In particular, X-ray mode switching has been used to discover new tMSP candidates in the disk state. In the pulsar state, features such as the X-ray luminosity ($\sim 10^{30}-10^{33}$~erg~s$^{-1}$), spectral shape (photon index $\simeq 1.5-3$), or orbital modulation from the IBS (often double peaked with a flux maximum near pulsar inferior conjunction) can aid in identifying spider systems.
\end{enumerate}

\subsection{Spider Definition and Taxonomy}
\label{sec:definition}

In our catalog, spider pulsars are defined as systems that meet the following three criteria:
\begin{enumerate}
    \item a pulsar spin period $P < 30$~ms (the majority of identified systems have spins $1.5<P<5.0$ ms);\footnote{This limit is a convention commonly used in the literature for MSPs (e.g., \citealt{Bhattacharya91,Lorimer08}) and corresponds to a sparsely populated region in the spin period distribution separating the main pulsar and MSP populations (see Section~\ref{sec:spins}).}
    \item an orbital period $P_{\rm b} \lesssim 10$~days (most systems have $P_{\rm b} \lesssim 1$~day, indicative of compact orbits; the exceptions are the so-called huntsman (HM) pulsars, defined below, which have wider orbits of 2--10~days); and
    \item a nondegenerate or semidegenerate companion star, inferred from (i) the presence of radio eclipses, (ii) the detection of X-ray mode switching (in tMSPs), (iii) the optical--IR spectra of the companion or accretion disk, and (iv) the detection and properties of X-ray and/or optical orbital modulation (see details above).
\end{enumerate}

Spiders are divided into two broad subclasses, based on the minimum companion mass ($M_{\rm c,min}$):
\begin{enumerate}
    \item RBs have low-mass main-sequence companions with $M_{c,\text{min}} \gtrsim 0.1 \, M_\odot$, and     
    \item BWs have ultralow-mass companions with $0.004 \, M_\odot < M_{c,\text{min}} < 0.1 \, M_\odot$. This group includes tidarrens \citep{romani16,Draghis19}---systems with $M_{c,\text{min}} < 0.02 \, M_\odot$ and $P_{\rm b}< 2$~hr---labeled ``ti'' below. However, we exclude pulsar-planet systems from our BW definition \citep[i.e., those with $M_{c,\text{min}} < 0.004 \, M_\odot$; see discussion in ][]{laycock25}.
\end{enumerate}

For completeness, we also include a related class of systems known in the literature as HMs:

\begin{enumerate}
\item HM pulsars are systems with giant companions and $P_{\rm b}$ in the 2--10~day range \citep{strader15, ssj+17, sru+25}. Due to their luminous companions and wide orbits, the companion stars are very weakly irradiated and are likely not ablated. However, these companions have masses similar to those found in RB systems and exhibit evidence of IBSs, akin to those seen in canonical spider systems \citep{swihart18}. 
\end{enumerate}

Thus, our definition includes three quantitative criteria ($P<30$~ms, $P_{\rm b} \lesssim 10$~days and $M_{\rm c,min}>0.004$~M$_\odot$) and one qualitative criterion assessing the nature of the companion star (nondegenerate or semidegenerate).
The latter is necessary to exclude MSP--WD systems with fully degenerate companions, which can overlap with spiders in $P$, $M_{\rm c,min}$, and sometimes $P_{\rm b}$.
As the sample grows---and as discussed below---peculiar systems are being discovered. Examples include the RB PSR~J1932+2121 with $P=14.2$~ms \citep{Misra25b} and the RB PSR~J1908+2105 with $M_{\rm c,min}=0.05$~M$_\odot$ \citep{simpson25}.

\section{Spider-Cat: A Catalog of Compact Binary Millisecond Pulsars} 

We present the spider catalog (SpiderCat), which compiles spin, orbital, and multiwavelength information for all publicly known spider systems in the Galactic field, excluding those in globular clusters.\footnote{For globular cluster MSPs, see \url{https://www3.mpifr-bonn.mpg.de/staff/pfreire/GCpsr.html}.} The paper is accompanied by an online database,\footnote{\url{https://astro.phys.ntnu.no/SpiderCAT}} where the information presented here is available for easy access.

In addition to confirmed spider systems, we also include spider candidates (denoted RB candidates (RBcs) and BW candidates (BWcs)), which exhibit similar characteristics to confirmed systems but lack definitive pulsation detections. These candidates are identified based on several observational signatures, as discussed in Section~\ref{sec:discoveries}. Although pulsations have not yet been detected in these systems, their multiwavelength properties strongly suggest a classification as spider MSPs.\footnote{Indeed, follow-up observations of several spider candidates have detected pulsations, e.g., in the BW PSR~J1311-3430 \citep{r12, pgf+12}, the RB PSR~J2339-0533 \citep{rs11,pc15} or the RB PSR~J0212+5321 \citep{Linares17,pbh+23}.}

In the following sections, we describe the source sample (Section~\ref{sec:sample}) and catalog parameters (Section~\ref{sec:params}). We present the Galactic distribution of spiders in Section~\ref{sec:distri}, the spider $P-\dot{P}$-diagram in Section~\ref{sec:spins}, and the $M_{\rm c,min}-P_{\rm b}$ diagram in Section~\ref{sec:orbits}. In Section~\ref{sec:distances}, we discuss and tabulate distance estimates for spider systems. Section~\ref{sec:counterparts} presents the multiwavelength counterparts ($\gamma$-ray, X-ray, optical, and IR), including their fluxes, estimated luminosities, and apparent and absolute magnitudes. We discuss potential spin-orbit grouping of spiders in Section~\ref{sec:spin-orbit} and conclude in Section~\ref{sec:conclusions}.  

\subsection{Sample}
\label{sec:sample}

%%%%%%%%%%%%%%%%%%%%%%%%%%%%
% Recap table
%%%%%%%%%%%%%%%%%%%%%%%%%%%%

\begin{deluxetable}{lccccc}
\label{tab:recap}
\tabletypesize{\scriptsize}
\caption{Number of Spiders by Type, along with the Number That Have Associated Multiwavelength Counterparts in the Catalogs Searched in This Paper and Are Also Listed in the ATNF Pulsar Catalog.}
\tablehead{& \colhead{No.} & \colhead{No. $\gamma$ ray} & \colhead{No. X-Ray} & \colhead{No. Optical} & \colhead{No. ATNF}
}
\startdata
    RB  & 30  & 23  & 21  & 25 & 25 \\
    BW  & 50  & 35  & 20  & 19 & 39 \\
    HM  & 2   & 2   & 1   & 2  & 1 \\
    RBc & 23  & 22  & 10  & 21 & 0 \\
    BWc & 5  & 4  & 2  &  5 & 0 \\
    HMc & 1  & 0  & 1  &  1 & 0 \\
    \hline
    Total & 111 & 86 & 55 & 73 & 65
\enddata
\end{deluxetable}

At the time of writing, SpiderCat includes 111 systems: 30 confirmed RBs, 50 confirmed BWs, two confirmed HMs, 23 RBcs, five BWcs, and one HM candidate (HMc; see Table~\ref{tab:recap}). 
We compiled all information from sources in the literature (as detailed in Table~\ref{tab:spidercat}) and public catalogs, including the ATNF Pulsar Catalog \citep{ATNFpsrcat}\footnote{\url{ http://www.atnf.csiro.au/research/pulsar/psrcat/}} and the Galactic MSP list \citep{GalMSPs},\footnote{\url{https://www.astro.umd.edu/~eferrara/GalacticMSPs.html}} as well as several catalogs of multiwavelength data (see Section~\ref{sec:counterparts} for details). 

Determining the precise discovery date of a spider system is often challenging, as many are found through radio surveys that require extensive postobservation analysis. To provide a consistent metric, we define the discovery year as the year in which a paper was published identifying the system as a spider or spider candidate, or reporting the detection of millisecond pulsations from the source. Figure~\ref{fig:disc_year} shows a histogram of spider discoveries by year.

As seen in Figure~\ref{fig:disc_year}, the first known spider system, PSR B1957$+$20, was discovered in 1988 \citep{fst88}, with only three additional Galactic-field systems identified in the pre-Fermi era. Two of these were BWs, PSR J2051$-$0827 \citep{sbl+96} and PSR J0610$-$2100 \citep{bjd+06}, while PSR J1723$-$2837 was discovered in 2004 \citep{fsk+04} and later classified as an RB \citep{cls+13}. The three prominent spikes in the histogram correspond to two major radio surveys targeting unidentified Fermi sources: one with the GBT \citep[six sources;][]{hrm+11}, one with the Arecibo Observatory \citep[five sources;][]{cck+16}, and the more recent FAST-GPPS \citep[13 sources;][]{why+24}.

%%%%%%%%%%%%%%%%%%%%%%%%%%%%
% SpiderCat
%%%%%%%%%%%%%%%%%%%%%%%%%%%%

\subsection{Catalog Parameters} \label{sec:params}

In Table~\ref{tab:spidercat}, we list all sources along with the following parameters: \\\\
(1) \textit{Name}---source name. ``PSR'' if confirmed pulsar. Otherwise the name used in the discovery paper.\\
(2) \textit{Type}---spider type: BW, RB, HM, BWc, RBc, and HMc. \\
(3) \textit{Sub.}---spider subtype: tr = tMSP, ti = tidarren. \\
(4) \textit{R.A.}---(J2000) (hh:mm:ss.ssss), where seconds are rounded to four digits. \\
(5) \textit{Decl.}---(J2000) (±dd:mm:ssss), where seconds are rounded to four digits. \\
(6) \textit{Year}---year of discovery publication. \\
(7) \textit{Ref. discovery}---reference for the discovery publication. \\
(8) \textit{$P$}---barycentric spin period of the pulsar (milliseconds). \\
(9) \textit{$\dot{E}$}---spin-down luminosity (in units of erg s$^{-1}$). \\
(10) \textit{$P_{\rm b}$}---binary period of pulsar (hours). \\
(11) \textit{$M_{\rm c, min}$}---the minimum companion mass ($M_{\odot}$). \\
(12) \textit{Ref. dyn.}---reference for the dynamical parameters (full timing solution when available).\\

We calculated the spin-down luminosity using the following equation: 

\begin{equation}
    \dot{E} =  4 \pi^2 I \frac{\dot{P}}{P^3},
    \label{eq:Edot}
\end{equation}
where we assume a ``canonical'' neutron star moment of inertia, $I = 2MR^2/5 \approx 10^{45}$ g cm$^{2}$, based on a neutron star mass $M_{\rm NS}=1.35 \, M_{\odot}$ and radius $R_{\rm NS}=10$ km. The tabulated $\dot{E}$ values are not corrected for the Shklovskii effect \citep{shklovskii70}.

Using information from radio timing, the ``mass function'' of the binary companion is given by:
\begin{equation}
    f_{\rm c} \equiv 
    \frac{(M_{\rm c}\sin i)^3}{(M_{\rm NS} + M_{\rm c})^2} =
    \frac{4\pi^2c^3}{GM_\odot}\frac{x_{\rm p}^3}{P_{\rm b}^2},
    \label{eq:massfunc}
\end{equation}
where $M_{\rm NS}$ is the mass of the pulsar, $M_{\rm c}$ is the mass of the companion, $i$ the orbital inclination, and $x_{\rm p} = a_{\rm p}\sin i/c$ is the projected semimajor axis of the pulsar's orbit.
Assuming $i=90^\circ$ and a ``canonical" neutron star mass ($M_{\rm NS}=1.35$~M$_\odot$) yields a lower limit on the companion mass, denoted $M_{\rm c,min}$.

In addition, Section~\ref{sec:counterparts} presents the X-ray, $\gamma$-ray and optical counterparts when identified (see Tables \ref{tab:xg_counterparts} and \ref{tab:oir_counterparts}), including the corresponding catalog source names and their angular separations from the spider coordinates listed in Table~\ref{tab:spidercat}. Table~\ref{tab:dist_lum} contains the derived source distances (described in detail in Section~\ref{sec:distances}), as well as X-ray and $\gamma$-ray fluxes and luminosities, $g$-band apparent and absolute magnitudes, and $g-r$ colors.

\startlongtable
\begin{deluxetable*}{lccccccccccc}  
\tabletypesize{\scriptsize}
\tablewidth{0pt}
\tablecaption{Spider Catalog \label{tab:spidercat}}
\tablehead{\colhead{Name} & \colhead{Type} & \colhead{Sub.} & \colhead{R.A.} & \colhead{Decl.} & \colhead{Year} & \colhead{Ref.} & \colhead{P} & \colhead{$\dot{E}$} & \colhead{P$_{\rm b}$} & \colhead{M$_{\rm c,min}$} & \colhead{Ref.} \\ \colhead{} & \colhead{} & \colhead{} & \colhead{(hh:mm:ss.ssss)} & \colhead{(dd:mm:ss.ssss)} & \colhead{} & \colhead{Discovery} & \colhead{(ms)} & \colhead{($10^{34}$ erg s$^{-1}$)} & \colhead{(hr)} & \colhead{(M$_{\odot}$)} & \colhead{Dyn.} \\ \colhead{(1)} & \colhead{(2)} & \colhead{(3)} & \colhead{(4)} & \colhead{(5)} & \colhead{(6)} & \colhead{(7)} & \colhead{(8)} & \colhead{(9)} & \colhead{(10)} & \colhead{(11)} & \colhead{(12)}} 
\startdata
\hline
 PSR J0023+0923        & BW  &  \ldots  & 00:23:16.8803 & +09:23:23.8902 & 2011 & 1  & 3.05  & 1.59 & 3.33   & 0.016 & 71 \\
 PSR J0251+2606        & BW  &  \ldots  & 02:51:2.5537  & +26:06:9.97    & 2016 & 2  & 2.54  & 1.82 & 4.86   & 0.024 & 72 \\
 PSR J0312-0921        & BW  &  \ldots  & 03:12:6.2147  & -09:21:56.5532 & 2021 & 3  & 3.70  & 1.53 & 2.34   & 0.009 & 16 \\
 PSR J0541+2959g       & BW  &  \ldots  & 05:41:44      & +29:59         & 2024 & 4  & 3.21  & \ldots     & 9.01   & 0.057 & 4  \\
 PSR J0610-2100        & BW  &  \ldots  & 06:10:13.5967 & -21:00:27.8991 & 2006 & 5  & 3.86  & 0.85 & 6.86   & 0.021 & 73 \\
 PSR J0636+5128        & BW  & ti & 06:36:4.8471  & +51:28:59.9658 & 2014 & 6  & 2.87  & 0.59 & 1.60   & 0.007 & 74 \\
 PSR J0952-0607        & BW  &  \ldots  & 09:52:8.3214  & -06:07:23.49   & 2017 & 7  & 1.41  & 6.67 & 6.42   & 0.019 & 75 \\
 PSR J1124-3653        & BW  &  \ldots  & 11:24:1.116   & -36:53:19.087  & 2011 & 1  & 2.41  & 1.7  & 5.45   & 0.027 & 76 \\
 PSR J1221-0633        & BW  &  \ldots  & 12:21:24.7582 & -06:33:51.6998 & 2023 & 8  & 1.93  & 2.87 & 9.27   & 0.013 & 8  \\
 PSR J1301+0833        & BW  &  \ldots  & 13:01:38.26   & +08:33:57.5    & 2012 & 9  & 1.84  & 6.65 & 6.54   & 0.024 & 16 \\
 PSR J1311-3430        & BW  & ti & 13:11:45.7242 & -34:30:30.35   & 2012 & 10 & 2.56  & 4.93 & 1.56   & 0.008 & 77 \\
 PSR J1317-0157        & BW  &  \ldots  & 13:17:40.4489 & -01:57:30.107  & 2023 & 8  & 2.91  & 0.88 & 2.14   & 0.018 & 8  \\
 PSR J1356+0230\tablenotemark{a}  & BW  & \ldots   & 13:56:37.2    & +02:30:28.8    & 2021 &  \ldots  & 2.83  & 1.37 & \ldots & \ldots & \ldots \\
 PSR J1446-4701        & BW  &  \ldots  & 14:46:35.7092 & -47:01:26.8138 & 2012 & 11 & 2.19  & 3.66 & 6.66   & 0.019 & 78 \\
 PSR J1513-2550        & BW  &  \ldots  & 15:13:23.3206 & -25:50:31.285  & 2016 & 12 & 2.12  & 8.97 & 4.29   & 0.016 & 12 \\
 PSR J1544+4937        & BW  &  \ldots  & 15:44:4.4911  & +49:37:55.374  & 2013 & 13 & 2.16  & 1.1  & 2.90   & 0.017 & 79 \\
 PSR J1555-2908        & BW  &  \ldots  & 15:55:40.6585 & -29:08:28.4232 & 2022 & 14 & 1.79  & 30.8 & 5.60   & 0.051 & 14 \\
 PSR J1602-1009        & BW  &  \ldots  & 16:02:13.2805 & -10:09:19.9199 & 2024 & 15 & 3.12  & 0.62 & 2.99   & 0.018 & 15 \\
 PSR J1627+3219        & BW  &  \ldots  & 16:27:52.9985 & +32:18:26.643  & 2023 & 16 & 2.18  & 2.08 & 3.98   & 0.022 & 16 \\
 PSR J1630+3550        & BW  &  \ldots  & 16:30:35.949  & +35:50:42.477  & 2022 & 17 & 3.23  & 2.64 & 7.58   & 0.01  & 54 \\
 PSR J1641+8049        & BW  &  \ldots  & 16:41:20.8331 & +80:49:52.9233 & 2018 & 18 & 2.02  & 4.68 & 2.18   & 0.04  & 80 \\
 PSR J1653-0158        & BW  & ti & 16:53:38.0538 & -01:58:36.893  & 2014 & 19 & 1.97  & 1.24 & 1.25   & 0.01  & 81 \\
 PSR J1705-1903        & BW  &  \ldots  & 17:05:43.8486 & -19:03:41.4172 & 2019 & 20 & 2.48  & \ldots     & 4.41   & 0.041 & 78 \\
 PSR J1720-0533        & BW  &  \ldots  & 17:20:54.5059 & -05:34:23.8223 & 2021 & 21 & 3.27  & 0.93 & 3.16   & 0.029 & 82 \\
 PSR J1731-1847        & BW  &  \ldots  & 17:31:17.6098 & -18:47:32.666  & 2011 & 22 & 2.34  & 7.78 & 7.47   & 0.033 & 83 \\
 PSR J1745-23          & BW  &  \ldots  & 17:45:30      & -23:25         & 2020 & 23 & 5.42  & \ldots     & 3.97   & 0.026 & 23 \\
 PSR J1745+1017        & BW  &  \ldots  & 17:45:33.8371 & +10:17:52.523  & 2013 & 24 & 2.65  & 0.58 & 17.53  & 0.014 & 24 \\
 PSR J1805+0615        & BW  &  \ldots  & 18:05:42.3997 & +06:15:18.606  & 2016 & 2  & 2.13  & 9.31 & 8.08   & 0.023 & 72 \\
 PSR J1810+1744        & BW  &  \ldots  & 18:10:37.2848 & +17:44:37.38   & 2011 & 1  & 1.66  & 3.84 & 3.56   & 0.043 & 76 \\
 PSR J1814+0045g       & BW  &  \ldots  & 18:14:10      & +00:45         & 2024 & 4  & 2.31  & \ldots     & 4.86   & 0.055 & 4  \\
 PSR J1830-0106g       & BW  &  \ldots  & 18:30:7       & -01:06         & 2024 & 4  & 1.76  & \ldots     & 2.56   & 0.046 & 4  \\
 PSR J1833-3840        & BW  &  \ldots  & 18:33:4.5823  & -38:40:46.072  & 2023 & 16 & 1.87  & 10.8 & 21.61  & 0.008 & 84 \\
 PSR J1838+1507g       & BW  &  \ldots  & 18:38:36      & +15:07         & 2024 & 4  & 3.82  & \ldots     & 2.69   & 0.028 & 4  \\
 PSR J1847+0342g       & BW  &  \ldots  & 18:47:30      & +03:42         & 2024 & 4  & 4.29  & \ldots     & 3.34   & 0.017 & 4  \\
 PSR J1919+0126g       & BW  &  \ldots  & 19:19:23      & +01:26         & 2024 & 4  & 1.90  & \ldots     & 6.47   & 0.025 & 4  \\
 PSR J1928+1245        & BW  &  \ldots  & 19:28:45.3936 & +12:45:53.374  & 2019 & 25 & 3.02  & 2.4  & 3.28   & 0.009 & 25 \\
 PSR J1946-5403        & BW  &  \ldots  & 19:46:34.497  & -54:03:42.51   & 2015 & 26 & 2.71  & \ldots     & 3.12   & 0.021 & 26 \\
 PSR J1953+1006g       & BW  &  \ldots  & 19:53:34      & +10:06         & 2024 & 4  & 2.59  & \ldots     & 2.89   & 0.012 & 4  \\
 PSR B1957+20          & BW  &  \ldots  & 19:59:36.7699 & +20:48:15.1222 & 1988 & 27 & 1.61  & 16.0 & 9.17   & 0.021 & 85 \\
 PSR J2003+3032g       & BW  &  \ldots  & 20:03:53      & +30:32         & 2024 & 4  & 1.79  & \ldots     & 3.59   & 0.03  & 4  \\
 PSR J2017-1614        & BW  &  \ldots  & 20:17:46.1478 & -16:14:15.51   & 2016 & 12 & 2.31  & 0.78 & 2.35   & 0.026 & 12 \\
 PSR J2047+1053        & BW  &  \ldots  & 20:47:10.246  & +10:53:7.8     & 2012 & 9  & 4.29  & 1.04 & 2.98   & 0.035 & 16 \\
 PSR J2051-0827        & BW  &  \ldots  & 20:51:7.5198  & -08:27:37.7497 & 1996 & 28 & 4.51  & 0.55 & 2.38   & 0.027 & 86 \\
 PSR J2052+1219        & BW  &  \ldots  & 20:52:47.778  & +12:19:59.022  & 2016 & 2  & 1.99  & 3.38 & 2.75   & 0.033 & 72 \\
 PSR J2055+3829        & BW  &  \ldots  & 20:55:10.3065 & +38:29:30.9057 & 2017 & 29 & 2.09  & 0.43 & 3.11   & 0.022 & 29 \\
 PSR J2115+5448        & BW  &  \ldots  & 21:15:11.7678 & +54:48:45.154  & 2016 & 12 & 2.60  & 16.8 & 3.25   & 0.021 & 12 \\
 PSR J2214+3000        & BW  &  \ldots  & 22:14:38.8524 & +30:00:38.2061 & 2011 & 30 & 3.12  & 1.92 & 10.00  & 0.013 & 71 \\
 PSR J2234+0944        & BW  &  \ldots  & 22:34:46.8558 & +09:44:30.0544 & 2012 & 9  & 3.63  & 1.66 & 10.07  & 0.015 & 87 \\
 PSR J2241-5236        & BW  &  \ldots  & 22:41:42.0402 & -52:36:36.2591 & 2011 & 31 & 2.19  & 2.6  & 3.50   & 0.012 & 88 \\
 PSR J2256-1024        & BW  &  \ldots  & 22:56:56.3929 & -10:24:34.385  & 2011 & 32 & 2.29  & 3.71 & 5.11   & 0.029 & 89 \\
 4FGL J0336.0+7502     & BWc &  \ldots  & 03:36:10.18   & +75:03:17.27   & 2021 & 33 & \ldots & \ldots & 3.72   & \ldots & \ldots   \\
 ZTF J1406+1222        & BWc &  \ldots  & 14:06:56.17   & +12:22:43.4    & 2022 & 34 & \ldots & \ldots & 1.03   & \ldots & \ldots   \\
 4FGL J1408.6-2917     & BWc &  \ldots  & 14:08:26.79   & -29:22:21.21   & 2022 & 35 & \ldots & \ldots & 3.42   & \ldots & \ldots   \\
 4FGL J1544.2-2554     & BWc &  \ldots  & 15:44:15.46   & -25:55:32.69   & 2025 & 36 & \ldots & \ldots & 2.72   & \ldots & \ldots   \\
 4FGL J1838.2+3223     & BWc &  \ldots  & 18:38:16.82   & +32:24:11.41   & 2024 & 37 & \ldots & \ldots & 4.02   & \ldots & \ldots   \\
 PSR J1417-4402        & HM  &  \ldots  & 14:17:30.6    & -44:02:57.4    & 2016 & 38 & 2.66  & \ldots     & 128.97 & 0.22  & 38 \\
 PSR J1947-1120        & HM  &  \ldots  & 19:47:38.24   & -11:20:27.21   & 2025 & 39 & 2.24  & \ldots     & 246.34 & 0.198 & 39 \\
 2FGL J0846.0+2820     & HMc &  \ldots  & 08:46:21.89   & +28:08:41      & 2017 & 40 & \ldots      &      & 195.19 & \ldots      &  \ldots  \\
 PSR J0212+5321        & RB  &  \ldots  & 02:12:10.47   & +53:21:38.81   & 2016 & 41 & 2.11  & \ldots     & 20.87  & 0.34  & 90 \\
 PSR J0838-2827        & RB  &  \ldots  & 08:38:50.4181 & -28:27:56.9729 & 2017 & 42 & 3.62  & 0.87 & 5.15   & 0.165 & 91 \\
 PSR J0955-3947        & RB  &  \ldots  & 09:55:27.8087 & -39:47:52.2931 & 2018 & 43 & 2.02  & 18.2 & 9.30   & 0.292 & 91 \\
 PSR J1023+0038        & RB  & tr & 10:23:47.6872 & +00:38:40.8455 & 2009 & 44 & 1.69  & 5.69 & 4.75   & 0.135 & 92 \\
 PSR J1036-4353        & RB  &  \ldots  & 10:36:30.2151 & -43:53:8.7252  & 2023 & 45 & 1.68  & 5.31 & 6.23   & 0.227 & 93 \\
 PSR J1048+2339        & RB  &  \ldots  & 10:48:43.4184 & +23:39:53.4043 & 2016 & 2  & 4.67  & 1.17 & 6.01   & 0.301 & 72 \\
 PSR J1227-4853        & RB  & tr & 12:27:58.724  & -48:53:42.741  & 2014 & 46 & 1.69  & 9.13 & 6.91   & 0.211 & 94 \\
 PSR J1242-4712        & RB  &  \ldots  & 12:42:12.7954 & -47:12:18.408  & 2024 & 47 & 5.31  & 0.56 & 7.77   & 0.087 & 47 \\
 PSR J1302-3258        & RB  &  \ldots  & 13:02:25.5262 & -32:58:36.843  & 2011 & 1  & 3.77  & 0.48 & 18.83  & 0.146 & 76 \\
 PSR J1306-4035        & RB  &  \ldots  & 13:06:56.3    & -40:35:23.3    & 2018 & 48 & 2.20  & \ldots     & 26.33  & 0.44  & 48 \\
 PSR J1431-4715        & RB  &  \ldots  & 14:31:44.6177 & -47:15:27.574  & 2015 & 49 & 2.01  & 6.84 & 10.79  & 0.124 & 49 \\
 PSR J1622-0315        & RB  &  \ldots  & 16:22:59.6285 & -03:15:37.328  & 2016 & 12 & 3.85  & 0.79 & 3.88   & 0.097 & 16 \\
 PSR J1628-3205        & RB  &  \ldots  & 16:28:7       & -32:05:48.9    & 2012 & 9  & 3.21  & 1.42 & 5.04   & 0.156 & 16 \\
 PSR J1723-2837        & RB  &  \ldots  & 17:23:23.1856 & -28:37:57.17   & 2004 & 50 & 1.86  & 4.66 & 14.77  & 0.236 & 95 \\
 PSR J1803-6707        & RB  &  \ldots  & 18:03:4.2353  & -67:07:36.1576 & 2023 & 45 & 2.13  & 7.48 & 9.13   & 0.288 & 93 \\
 PSR J1816+4510        & RB  &  \ldots  & 18:16:35.9346 & +45:10:33.855  & 2012 & 9  & 3.19  & 5.22 & 8.66   & 0.158 & 74 \\
 PSR J1849+0304g       & RB  &  \ldots  & 18:49:32      & +03:04         & 2024 & 4  & 1.79  & \ldots     & 5.79   & 0.279 & 4  \\
 PSR J1859+0313g       & RB  &  \ldots  & 18:59:36      & +03:13         & 2024 & 4  & 1.61  & \ldots     & 10.66  & 0.288 & 4  \\
 PSR J1908+2105        & RB  &  \ldots  & 19:08:57.294  & +21:05:2.726   & 2016 & 2  & 2.56  & 3.24 & 3.51   & 0.054 & 72 \\
 PSR J1910-5320        & RB  &  \ldots  & 19:10:49.1205 & -53:20:57.1205 & 2023 & 51 & 2.33  & 11.5 & 8.36   & 0.277 & 96 \\
 PSR J1919+1502g       & RB  &  \ldots  & 19:19:57      & +15:02         & 2024 & 4  & 3.65  & \ldots     & 2.69   & 0.095 & 4  \\
 PSR J1931+1428g       & RB  &  \ldots  & 19:31:48      & +14:28         & 2024 & 4  & 2.61  & \ldots     & 4.36   & 0.248 & 4  \\
 PSR J1932+2121        & RB  &  \ldots  & 19:32:21.201  & +21:21:6.78    & 2024 & 4  & 14.24 & 0.48 & 1.94   & 0.115 & 4  \\
 PSR J1957+2516        & RB  &  \ldots  & 19:57:34.6115 & +25:16:2.076   & 2015 & 52 & 3.96  & 1.74 & 5.72   & 0.097 & 97 \\
 PSR J2039-5617        & RB  &  \ldots  & 20:39:34.9681 & -56:17:9.268   & 2015 & 53 & 2.65  & 3.0  & 5.47   & 0.171 & 98 \\
 PSR J2055+1545        & RB  &  \ldots  & 20:55:47.8317 & +15:45:21.2164 & 2023 & 54 & 2.16  & 3.83 & 4.82   & 0.244 & 54 \\
 PSR J2129-0429        & RB  &  \ldots  & 21:29:45.05   & -04:29:6.81    & 2011 & 1  & 7.61  & 2.93 & 15.25  & 0.368 & 76 \\
 PSR J2215+5135        & RB  &  \ldots  & 22:15:32.6    & +51:35:36.3    & 2011 & 1  & 2.61  & 7.41 & 4.14   & 0.208 & 76 \\
 PSR J2333-5526        & RB  &  \ldots  & 23:33:15.9676 & -55:26:21.1056 & 2020 & 55 & 2.10  & 3.31 & 6.90   & 0.293 & 91 \\
 PSR J2339-0533        & RB  &  \ldots  & 23:39:38.74   & -05:33:5.32    & 2011 & 56 & 2.88  & 2.32 & 4.63   & 0.257 & 99 \\
 4FGL J0327.3+2355     & RBc &  \ldots  & 03:27:43.03   & +23:51:43.91   & 2025 & 57 & \ldots & \ldots & 6.98   & \ldots & \ldots \\
 4FGL J0407.7-5702     & RBc & tr & 04:07:31.72   & -57:00:25.3    & 2020 & 58 & \ldots & \ldots & \ldots & \ldots & \ldots \\
 3FGL J0427.9-6704     & RBc & tr & 04:27:49.61   & -67:04:35      & 2016 & 59 & \ldots & \ldots & 8.80   & 0.324 & \ldots \\
 1FGL J0523.5-2529     & RBc &  \ldots  & 05:23:16.93   & -25:27:36.9    & 2014 & 60 & \ldots & \ldots & 16.52  & \ldots & \ldots \\
 4FGL J0540.0-7552     & RBc & tr & 05:40:1.89    & -75:54:19.26   & 2021 & 61 & \ldots & \ldots & \ldots & \ldots & \ldots \\
 4FGL J0639.1-8009     & RBc & tr & 06:40:59.5    & -80:11:26.19   & 2025 & 62 & \ldots & \ldots & \ldots & \ldots & \ldots \\
 4FGL J0705.8-0004     & RBc &  \ldots  & 07:05:52.88   & +00:00:23.78   & 2025 & 57 & \ldots & \ldots & 7.49   & \ldots & \ldots \\
 3FGL J0737.2-3233     & RBc &  \ldots  & 07:36:56.22   & -32:32:55.3    & 2024 & 63 & \ldots & \ldots & 8.52   & \ldots & \ldots \\
 4FGL J0935.3+0901     & RBc &  \ldots  & 09:35:20.719  & +09:00:35.9    & 2020 & 64 & \ldots & \ldots & 2.44   & \ldots & \ldots \\
 4FGL J0940.3-7610     & RBc &  \ldots  & 09:40:23.79   & -76:10:0.1     & 2021 & 65 & \ldots & \ldots & 6.50   & \ldots & \ldots \\
 CXOU J110926.4-650224 & RBc & tr & 11:09:26.43   & -65:02:25      & 2019 & 66 & \ldots & \ldots & \ldots & \ldots & \ldots \\
 3FGL J1544.6-1125     & RBc & tr & 15:44:39.38   & -11:28:4.3     & 2015 & 67 & \ldots & \ldots & 5.80   & \ldots & \ldots \\
 4FGL J1646.5-4406     & RBc &  \ldots  & 16:46:22.75   & -44:05:41      & 2024 & 68 & \ldots & \ldots & 5.27   & \ldots & \ldots \\
 4FGL J1701.8-2226     & RBc &  \ldots  & 17:01:47.5    & -22:31:26.14   & 2025 & 57 & \ldots & \ldots & 7.80   & \ldots & \ldots \\
 4FGL J1702.7-5655     & RBc & tr & 17:02:51.01   & -56:55:9.1     & 2022 & 69 & \ldots & \ldots & 5.85   & \ldots & \ldots \\
 4FGL J1813.5+2819     & RBc &  \ldots  & 18:13:26.28   & +28:20:7.91    & 2025 & 57 & \ldots & \ldots & 7.05   & \ldots & \ldots \\
 4FGL J1819.4-1102     & RBc &  \ldots  & 18:19:48.52   & -11:03:41.88   & 2025 & 57 & \ldots & \ldots& 6.56   & \ldots & \ldots \\
 4FGL J1824.2+1231     & RBc & tr & 18:24:8.88    & +12:32:33.31   & 2025 & 62 & \ldots & \ldots&\ldots&\ldots& \ldots   \\
 4FGL J1853.6-0620     & RBc &  \ldots  & 18:53:23.66   & -06:18:40.89   & 2025 & 57 & \ldots & \ldots& 14.86  & \ldots & \ldots   \\
 4FGL J1859.2-0706     & RBc &  \ldots  & 18:59:18.57   & -07:10:1.1     & 2025 & 57 & \ldots & \ldots& 16.99  & \ldots &  \ldots  \\
 4FGL J1901.8-0718     & RBc &  \ldots  & 19:01:56.11   & -07:16:50.97   & 2025 & 57 & \ldots & \ldots& 8.37   & \ldots & \ldots   \\
 4FGL J2054.2+6904     & RBc &  \ldots  & 20:53:58.99   & +69:05:19.71   & 2023 & 70 & \ldots & \ldots& 14.93  & \ldots & \ldots   \\
 3FGL J2221.6+6507     & RBc &  \ldots  & 22:22:32.8    & +65:00:21      & 2024 & 63 & \ldots & \ldots& 3.96   & \ldots & \ldots   \\
\hline
\enddata
\tablerefs{(1) \citet{hrm+11};
(2) \citet{cck+16};
(3) \citet{trr+21};
(4) \citet{why+24};
(5) \citet{bjd+06};
(6) \citet{slr+14};
(7) \citet{bph+17};
(8) \citet{spp+23};
(9) \citet{rap+12};
(10) \citet{r12};
(11) \citet{kjb+12};
(12) \citet{san16};
(13) \citet{brr+13};
(14) \citet{rnc+22};
(15) \citet{msk+24};
(16) \citet{saa+23};
(17) \citet{sbo+22};
(18) \citet{lsk+18};
(19) \citet{rfc+14};
(20) \citet{mbc+19};
(21) \citet{www+21};
(22) \citet{bbb+11};
(23) \citet{ccb+20};
(24) \citet{bgc+13};
(25) \citet{pkr+19};
(26) \citet{ckr+15};
(27) \citet{fst88};
(28) \citet{sbl+96};
(29) \citet{goc+19};
(30) \citet{rrc+11};
(31) \citet{kjr+11};
(32) \citet{blm+11};
(33) \citet{ljhk21};
(34) \citet{bmf+22};
(35) \citet{ssc+22};
(36) \citet{kzz+25};
(37) \citet{zkk+24};
(38) \citet{crr+16};
(39) \citet{sru+25};
(40) \citet{ssj+17};
(41) \citet{lkh+16};
(42) \citet{hbt17};
(43) \citet{lhs+18};
(44) \citet{asr+09};
(45) \citet{cbb+23};
(46) \citet{bph+14};
(47) \citet{gbl+24};
(48) \citet{kbj+18};
(49) \citet{btb+15};
(50) \citet{fsk+04};
(51) \citet{ass+23};
(52) \citet{lbh+15};
(53) \citet{r15};
(54) \citet{lod+23};
(55) \citet{ssu+20};
(56) \citet{rs11};
(57) \citet{lrl+25};
(58) \citet{mss+20};
(59) \citet{slc+16};
(60) \citet{scs+14};
(61) \citet{ssu+21};
(62) \citet{krs+25};
(63) \citet{tlk+24};
(64) \citet{wxz+20};
(65) \citet{ssa+21};
(66) \citet{cpd+19};
(67) \citet{bh15};
(68) \citet{zwl+24};
(69) \citet{ccc+22};
(70) \citet{kzs+23};
(71) \citet{aab+21a};
(72) \citet{drc+21};
(73) \citet{vbc+22};
(74) \citet{flm+23};
(75) \citet{ncb+19};
(76) \citet{bbc+24};
(77) \citet{pgf+12};
(78) \citet{sbf+24};
(79) \citet{kbkr23};
(80) \citet{kzk+24};
(81) \citet{nck+20};
(82) \citet{mbz+23};
(83) \citet{nbb+14};
(84) \citet{kjc+25};
(85) \citet{aft94};
(86) \citet{svf+16};
(87) \citet{aab+21b};
(88) \citet{rsc+21};
(89) \citet{csm+20};
(90) \citet{pbh+23};
(91) \citet{tcb+24};
(92) \citet{akh+13};
(93) \citet{bnc+24};
(94) \citet{rrb+15};
(95) \citet{cls+13};
(96) \citet{dbc+24};
(97) \citet{sab+16};
(98) \citet{cnv+21};
(99) \citet{pc15}}
\tablenotetext{}{{\bf Notes.} See Section~\ref{sec:params} for details. (This table is available in machine-readable form in the online article.)} 
\tablenotetext{a}{This pulsar was discovered by TRAPUM's Fermi Sources working group using MeerKAT, targeting the sky position of 4FGL J1356.6+0234. The pulsar has been timed in $\gamma$-rays by Paolo Freire; \url{https://confluence.slac.stanford.edu/spaces/GLAMCOG/pages/383919352/Gamma-ray+PSR+J1356p0230}.}
\end{deluxetable*}

\section{Results and Discussion}

In this section, we present key statistical and physical properties of the 109 compact binary millisecond pulsars compiled in SpiderCat, focusing on their Galactic distribution, spin behavior, orbital characteristics, companion masses, and distances. Through multiwavelength data and population-wide trends, we provide insight into the demographics, evolution, and spatial distribution of spider MSPs. The following subsections highlight our main findings derived from the catalog, but independent analysis using the public SpiderCat database may uncover more emerging properties.

\subsection{Galactic Distribution} \label{sec:distri}

\begin{figure*}
\plotone{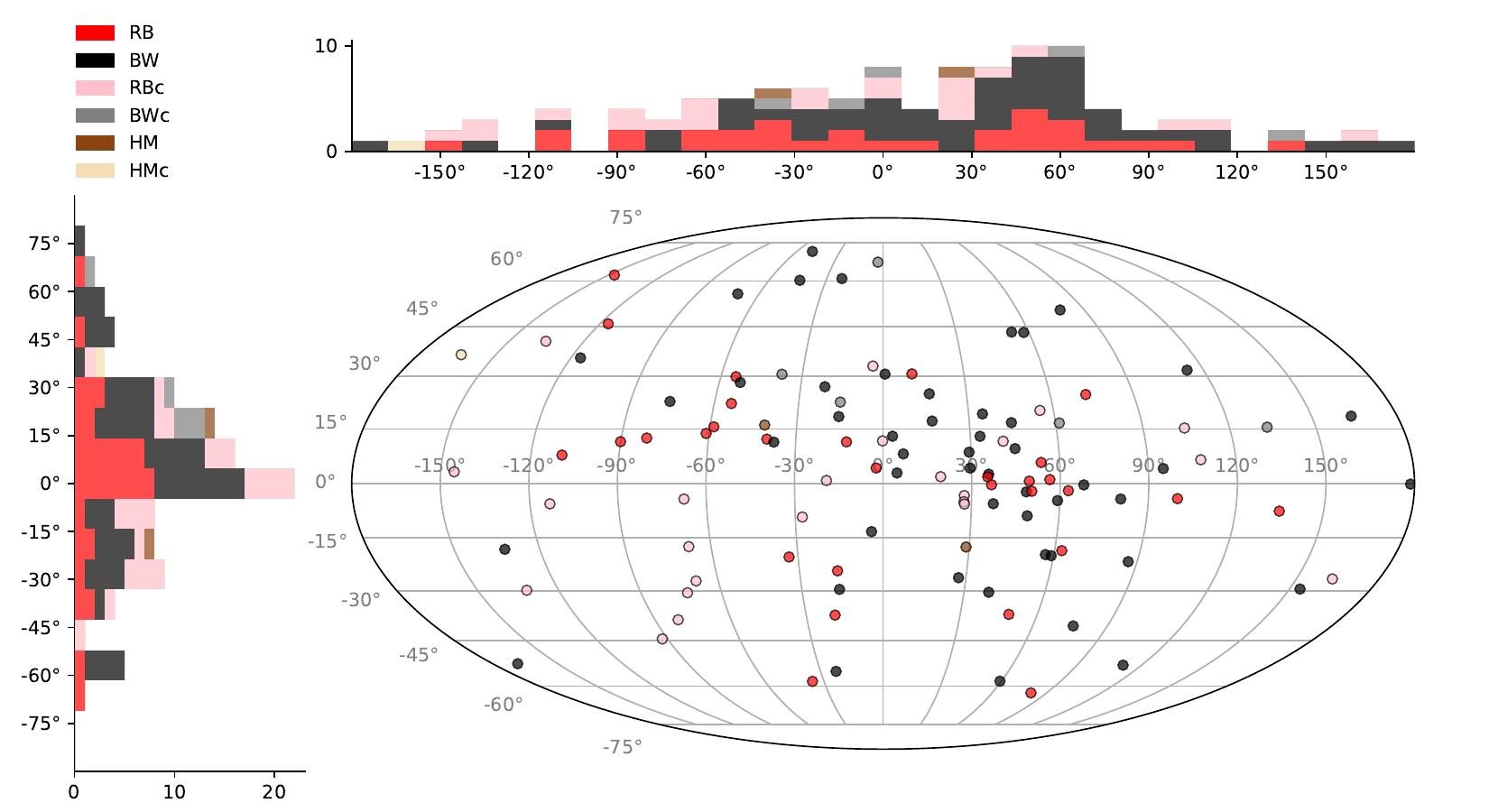}
\caption{Galactic distribution of spider systems in a Mollweide projection, color coded by spider subclass. The top and left panels show histograms of Galactic longitude (bin size: 12$^{\circ}$.4) and latitude (bin size: 9$^{\circ}$.5), respectively. The spider population is strongly concentrated toward low Galactic latitudes, while the distribution in Galactic longitude is relatively uniform.
\label{fig:skymap}}
\end{figure*}

\begin{figure}
\plotone{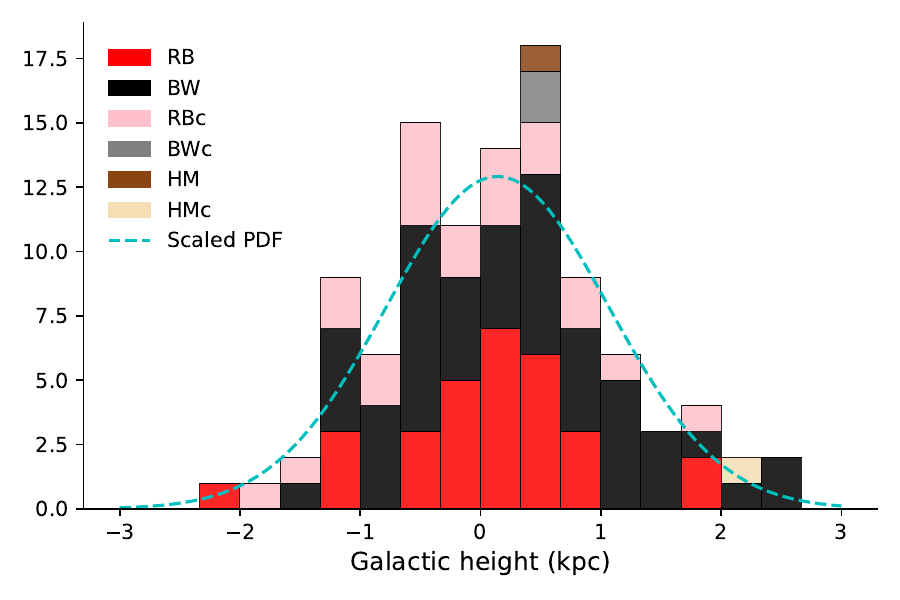}
\caption{Histogram of estimated Galactic heights for spider systems (bin size: 0.33 kpc), color coded by spider subclass. The distribution is approximately normal, with a mean height of $\bar{z} = 0.14 \pm 0.09$ kpc and a standard deviation of $\sigma_z = 0.93 \pm 0.06$ kpc, overplotted as a scaled Gaussian (dashed line). 
\label{fig:galactic_height}}
\end{figure}

The Galactic distribution of spiders is shown in Figure~\ref{fig:skymap}. Their Galactic latitude distribution is concentrated toward low latitudes, with the majority of sources located at $|b|<30^{\circ}$ (73\%), including 27 sources (25\%) within $|b|<6^{\circ}$. The distribution in Galactic longitude is relatively uniform, with most sources (68\%) found at $|l|<70^{\circ}$. There appears to be a concentration between $20^{\circ} < l < 70^{\circ}$, where 35 sources (32\%) are located. However, we note that this region includes 12 sources from FAST-GPPS, which may introduce an observational bias.
Most spider searches to date have avoided the Galactic plane ($|b|>5^{\circ}$), to minimize $\gamma$-ray diffuse emission and/or reddening.
On the other hand, the intrinsic number density of spiders (like that of most stellar populations) is expected to increase when approaching the Galactic plane.

The estimated Galactic heights, defined as $z = d\sin{b}$, where $d$ is the distance to the source, follow an approximately normal distribution, with a mean of $\bar{z} =0.14 \pm 0.09$ kpc and a standard deviation of $\sigma_{z} = 0.93 \pm 0.06$ kpc (Figure~\ref{fig:galactic_height}). These uncertainties are derived by sampling the distance distributions of all sources with distance measurements, as discussed in Section~\ref{sec:distances}. We also fit the histogram of the Galactic heights larger than $|z|> 0.45$ kpc with an exponential function to estimate the Galactic scale height ($z_e$) of spiders. This is to take into account the selection effect near the Galactic plane, where spiders have been searched less. Sampling the distance distribution resulted in $z_e = 0.72 \pm 0.14$ kpc, which is higher than the previous estimate \citep[$0.4\pm0.1$ kpc;][]{linares21}, at the 1.8$\sigma$ level. Our $z_e$ measurement remains consistent with estimates based on $\gamma$-ray-emitting MSPs \citep[][$z_e \sim 1.0$ kpc assuming known 1FGL MSPs, or $z_e \sim 0.6$ kpc taking into account pulsar-like unassociated Fermi sources]{gregoire13}, and broadly aligns with models of the Galactic MSP population \citep{lorimer13, levin13}. The current SpiderCat sample might indicate that spiders have a higher scale height than other MSPs, but a detailed investigation requires careful analysis of the selection effects and is beyond the scope of this work.
\subsection{Spin Periods} \label{sec:spins}

\begin{figure*}
\plotone{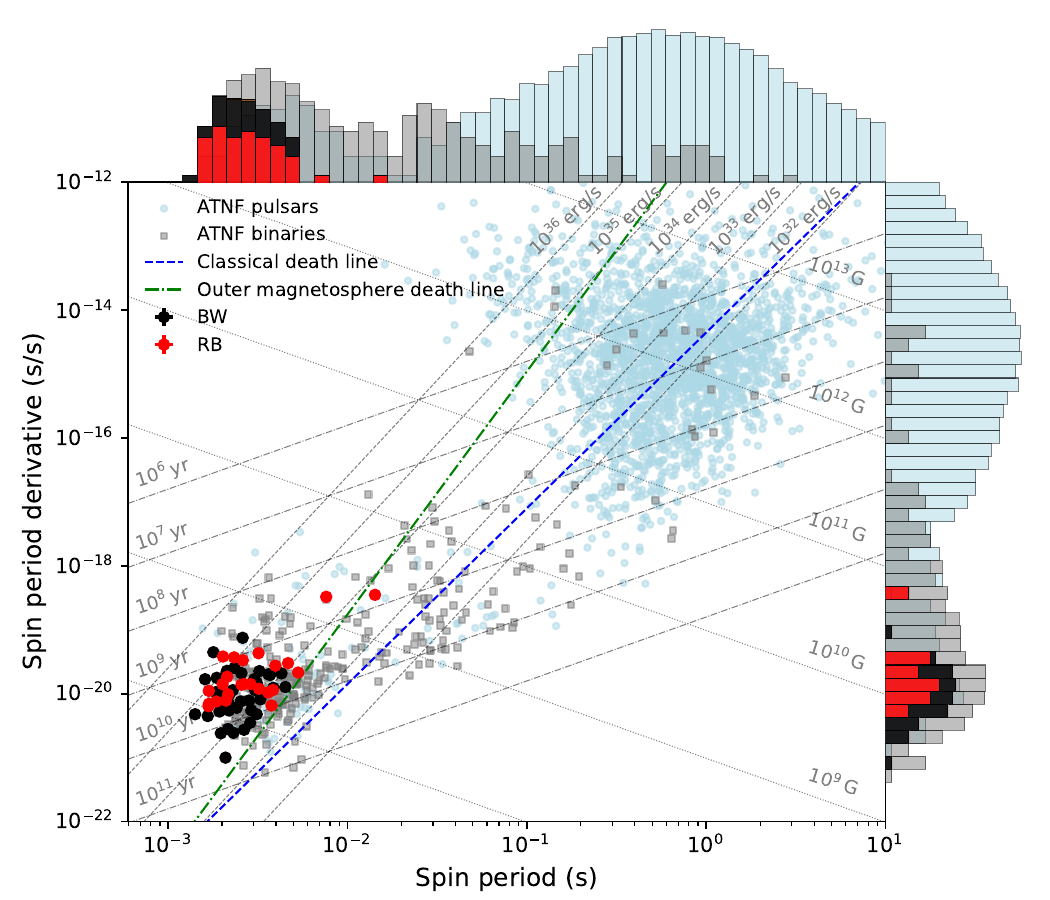}
\caption{Spin period ($P$) vs. spin period derivative ($\dot{P}$) diagram for ATNF pulsars, highlighting binaries and confirmed spider systems. Lines of constant $\dot{E}$, $B$ and $\tau$ are superposed. Also, the classical and outer magnetosphere ``death lines'' are shown as dashed blue and dashed-dotted green lines, respectively (see text). Histograms of $P$ and $\dot{P}$ are shown above and to the right of the plot, respectively, on a logarithmic scale for clarity. Spiders occupy the lower-left region of the $P/\dot{P}$-diagram, corresponding to the fastest-spinning, most weakly magnetized, and oldest pulsars. Two outlier systems, PSRs J2129$-$0429 and J1932$+$2121, lie above the main population, likely representing mildly recycled systems with longer spin periods and higher $\dot{P}$, indicative of alternative evolutionary paths.
\label{fig:ppdot}}
\end{figure*}

Pulsars found in spider systems are among the fastest-spinning MSPs, including the record holder BW PSR J0952$-$0607 \citep{ncb+19}, which has a spin period of just 1.41~ms (the fastest after the globular cluster RB PSR~J1748–2446ad, \citealt{Hessels06}). In the $P-\dot{P}$-diagram (Figure~\ref{fig:ppdot}), spiders occupy the lower-left corner at the end of the recycled pulsar branch and are characterized by short spin periods (median of 2.5~ms, with nearly all systems falling in the narrow 1.5--5.0~ms range) and spin period derivatives around $\dot{P}\approx10^{-20}$~s s$^{-1}$. As a result, their characteristic ages ($\tau\sim P/2\dot{P}$) are between $10^{9}$ and $10^{10}$~yr. Their characteristic surface magnetic fields ($B \propto \sqrt{P\dot{P}}$) are among the lowest inferred for pulsars, between $10^{8}$ and $10^{9}$~G. The spin-down luminosities ($\dot{E} \propto \dot{P}/P^3$) of spider systems range from $4\times10^{33}$~erg s$^{-1}$ to $3\times10^{35}$~erg s$^{-1}$ (see Table~\ref{tab:spidercat}).

We also show two theoretical ``death lines'' in Figure~\ref{fig:ppdot}, which represent approximate thresholds below which pair production, and hence pulsar activity, is expected to cease \citep{sturrock71, ruderman75, chen93}. The classical death line, derived under the polar cap particle acceleration model in a dipole magnetic field, is shown as a blue dashed line and corresponds to\footnote{Here we have converted the original relation between surface magnetic field strength and spin period into a relation between spin period derivative and spin period by assuming $B[\rm{G}] = \sqrt{\frac{3 c^3 I \dot{P} P}{8 \pi^2 R^6}} \approx 3.2 \times 10^{19} \sqrt{P\dot{P}}$, where $I = 10^{45}$ g cm$^2$ and $R=10$ km.} $\log{\dot{P}} = 2.75 \times P - 14.35$ (corresponding to Equation (6) in \citealt{chen93}). 

For an outer magnetosphere accelerator model, which is relevant for pulsars with high spin-down power \citep{cheng86a,cheng86b}, an alternative death line marking the cessation of pair production and strong $\gamma$-ray emission in the outer magnetosphere is shown as a green dash-dotted line and corresponds to $\log{\dot{P}} = 3.8 \times P - 11.16$ (Equation (27) in \citealt{chen93}). All spider systems lie above both death lines, with the exception of PSR J1932$+$2121, suggesting it may be near the limit for sustained $\gamma$-ray production.

Two notable outliers in the $P-\dot{P}$-diagram are PSR J2129$-$0429 ($P=7.5$ ms, $\dot{P}=3\times10^{-19}$ s s$^{-1}$) and PSR J1932$+$2121 ($P=14.2$ ms, $\dot{P}=3.5\times10^{-19}$ s s$^{-1}$), both of which have longer $P$ and higher $\dot{P}$ than the bulk of spiders. Notably, both are classified as RBs, which may have spent less time in the spider phase compared to typical BWs. This may account for their relatively mild recycling, potentially resulting from an inefficient mass-transfer phase or an early evolutionary stage, with the possibility of further spin-up during a subsequent Roche-Lobe overflow phase \citep{Misra25b}.

A limited amount of accretion would, in turn, imply low pulsar masses. Indeed, evolutionary calculations presented in \citet{Misra25b} support neutron star masses of $M_{\rm NS}\approx 1.6 \, M_{\odot}$. While no mass estimate currently exists for PSR J1932$+$2121, \citet{Clark23} has constrained the mass of PSR J2129$-$0429 to $1.61 \, M_{\odot} < M_{\rm NS} < 1.88 \, M_{\odot}$ (95\% confidence). 

\subsection{Orbits and Companion Masses} \label{sec:orbits}

\begin{figure*}
\plotone{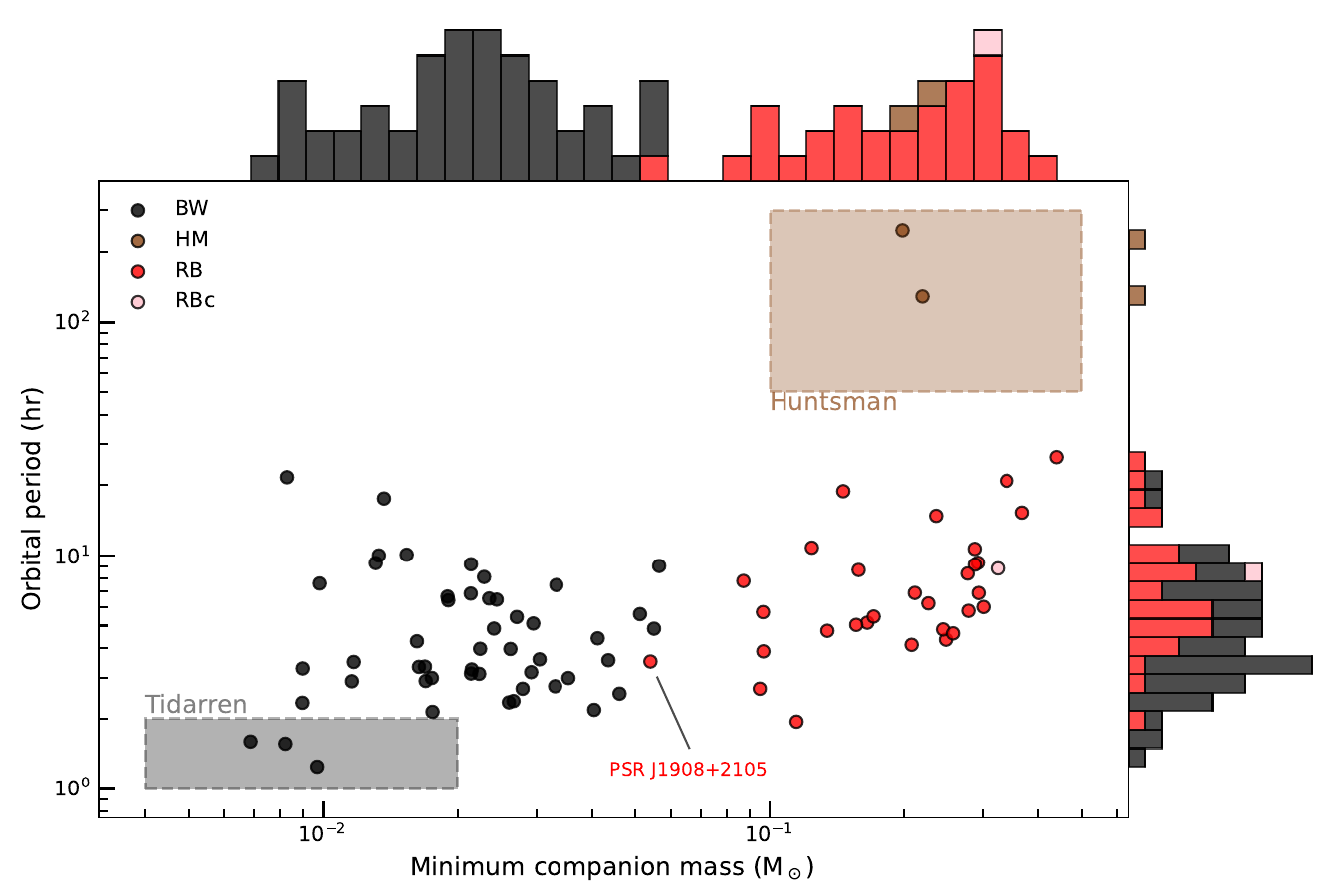}
\caption{Minimum companion mass vs. orbital period for spider systems, color coded by spider subclass. RBs and BWs are clearly separated by their characteristic companion masses, with RBs generally having $M_{c,\text{min}} \gtrsim 0.1 \, M_{\odot}$ and BWs $M_{c,\text{min}} \lesssim 0.06 \, M_{\odot}$. The areas of tidarren and HM systems are highlighted at the low-period and low-mass and high-period and high-mass corners, respectively. PSR J1908$+$2105 is labeled as an outlier RB with an anomalously low minimum companion mass, explained by a low orbital inclination.
\label{fig:mass_orbit}}
\end{figure*}

The minimum companion masses of spider systems range from 0.006 to 0.5~$M_{\odot}$, exhibiting a strongly bimodal distribution (Figure~\ref{fig:mass_orbit}). The highest minimum mass observed among BWs is approximately 0.06~$M_{\odot}$, while the minimum for RBs is around 0.1~$M_{\odot}$, with the exception of PSR J1908$+$2105. For this system, \citet{simpson25} found a high mass ratio ($q>0.55$) and companion temperature (about 5600~K), confirming its RB classification. 
They also derived a very low orbital inclination ($i<6^\circ$), explaining the unusually low minimum mass, since $M_{c,\text{min}}$ is calculated assuming $i=90^\circ$.
Extreme outliers aside, the minimum companion mass remains the most reliable parameter to distinguish between BWs and RBs.

Excluding the two known HM systems and the three tidarrens, the orbital periods of all other spiders are between 2 and 30~hr. While the difference in orbital periods between BWs and RBs is less distinct than that of companion masses, their population medians differ at the 2$\sigma$ level (Mood's median test: $\chi^2(1)=3.96$, p-value = 0.047). The median orbital period for RBs is $\bar{P}_{\rm b, RB} = 6.2$~hr (interquartile range: 4.8--10.8~hr), whereas for BWs it is $\bar{P}_{\rm b, BW} = 3.6$~hr (interquartile range: 2.9--6.6~hr). Only four RBs have orbital periods shorter than 4~hr (PSR~J1622$-$0315, PSR~J1908$+$2105, PSR~J1919$+$1502g, and PSR~J1932$+$2121), compared to 27 BWs in this regime.
We stress, however, that orbital period alone cannot be used to distinguish between RBs and BWs, as illustrated by the recently discovered RB PSR~J1932+2121 with a short $P_{\rm b}=1.94$~hr \citep{why+24,Misra25b}.

\subsection{Distances} \label{sec:distances}

Measuring distances to spider systems is challenging, as they are typically inferred from the DM of the radio pulsar and require models of the free electron density along the line of sight. In this paper, we estimate distances to spiders either via the DM-based method with the Galactic electron density model of \citet[][``YMW16'']{yao17} where available,\footnote{Using the \textsc{pygedm} package \citep{price21}.} or, alternatively, through geometric distances derived from Gaia DR3 parallaxes. In cases where the relative parallax uncertainty ($\sigma_\varpi / \varpi$) is less than 0.2, we always adopt the parallax-based distance, regardless of whether a DM-based estimate is available. In deriving the parallax-based distance we adopt the prior from \citet{bailerjones21}, and use the parallax-distance estimate even in cases where the parallax error-to-signal ratio exceeds one (when the distance posterior is heavily influenced by the prior). See also \citet[][Appendix A]{koljonen23} for details on distance estimates based on Gaia DR3 parallaxes.

In the following, we list exceptions to the above procedure:
\begin{enumerate}
\item {\it PSR J1720-0533}. We adopt the DM distance using the Galactic electron density model of \citet[][``NE2001'']{cordes02}. As discussed in \citet{koljonen24}, the inclusion of a dense component toward the North Polar Spur in the YMW16 model underestimates the distance to this BW, as revealed by flux upper limits on its IR counterpart. 
\item {\it PSR J0955-3947 and PSR J1036-4353}. We use NE2001 DM distances, which are more consistent with Gaia parallax estimates (see comparison in \citealt{koljonen23}). 
\item {\it PSR J1317-0157}. We adopt the NE2001 DM distance, as the YMW16 model yields a pegged and likely unphysical distance of 25 kpc.  
\item {\it PSR J0838-2827}. We use the distance derived from Gaia parallax measurements (see discussion in \citealt{tcb+24}). 
\item {\it PSR J0636+5128}. We rely on the NE2001 DM distance (see discussion in \citealt{draghis18}).
\end{enumerate}

When compared to parallax-based distances, DM-derived values tend to be systematically lower, by approximately 40\% \citep{jennings18, koljonen23}. Therefore, we conservatively adopt a relative uncertainty of $\delta d/d = 0.4$ (1$\sigma$) for DM-based distances. For the Gaia parallax-based distances, we use the full posterior distribution and define the 1$\sigma$ confidence interval as the 0.159 and 0.841 quantiles of the posterior. We note that, in most cases, the posterior distance distributions are asymmetric (see Table~\ref{tab:dist_lum}).

\subsection{Multiwavelength Counterparts} \label{sec:counterparts}

We first searched for matches within 1$\arcmin$ of our spider locations in the ATNF Pulsar Catalog \citep[v.2.6.0;][]{ATNFpsrcat}, using \textsc{psrqpy} \citep[v.1.3.2;][]{psrqpy}.
We used the following observational catalogs to search for associated multiwavelength counterparts obtained through the VizieR Service for Astronomical Catalogues \citep{ochsenbein00} to the spiders listed in Table~\ref{tab:spidercat}: Gaia DR3 \citep{gaia16b,gaia23}, the Two Micron All Sky Survey \citep[2MASS;][]{skrutskie06}, Panoramic Survey Telescope and Rapid Response System (Pan-STARRS) Data Release 2 \citep[PS1-DR2;][]{chambers16}, SkyMapper Southern Sky Survey (SMSS) Data Release 4 \citep[DR4;][]{onken24}, the Swift/XRT Point Source Catalogue \citep[2SXPS;][]{evans20}, Chandra Source Catalog Release 2.1 \citep[CSC2.1;][]{evans24}, eROSITA-DE Data Release 1 \citep[eRASS1;][]{merloni24}, the XMM-Newton Serendipitous Source Catalogue Data Release 14 \citep[4XMM-DR14;][]{webb20} and the Fermi-LAT 14 yr Source Catalog \citep[4FGL-DR4;][]{aab+22, ballet23}. We applied a search radius of 2$\arcsec$ for optical and IR catalogs, 2$\arcsec$--9$\arcsec$ for X-ray catalogs, and 10$\arcmin$ for the $\gamma$-ray catalog, centered on the positions (R.A. and decl.) listed in Table~\ref{tab:spidercat}. Source IDs of the identified counterparts are presented in Tables \ref{tab:xg_counterparts} and \ref{tab:oir_counterparts}, along with their angular separations from the query coordinates. Table~\ref{tab:dist_lum} includes distance estimates, X-ray and $\gamma$-ray fluxes and luminosities, as well as extinction-corrected $g$-band apparent and absolute magnitudes, and $g–r$ color indices of the counterparts.

\subsubsection{X-Rays and $\gamma$-Rays} \label{sec:xg}

\startlongtable
\begin{deluxetable*}{lcccccccccc}  
\tabletypesize{\scriptsize}
\tablewidth{0pt}
\tablecaption{X-Ray and $\gamma$-Ray Counterparts \label{tab:xg_counterparts}}
\tablehead{\colhead{Name} & \colhead{Fermi} & \colhead{Sep.} & \colhead{XMM-Newton} & \colhead{Sep.} & \colhead{Swift} & \colhead{Sep.} & \colhead{Chandra} & \colhead{Sep.} & \colhead{eROSITA} & \colhead{Sep.} \\ \colhead{} & \colhead{4FGL} & \colhead{($\arcmin$)} & \colhead{4XMM} & \colhead{($\arcsec$)} & \colhead{2SXPS} & \colhead{($\arcsec$)} & \colhead{2CXO} & \colhead{($\arcsec$)} & \colhead{1eRASS} & \colhead{($\arcsec$)}}
\startdata
\hline
 PSR J0023+0923        & J0023.4+0920  & 3.1 & \ldots & \ldots & \ldots & \ldots & J002316.8+092323 & 0.12 & \ldots & \ldots \\
 PSR J0212+5321        & J0212.1+5321  & 0.6 & \ldots & \ldots & 159668 & 0.6 & J021210.4+532138 & 0.04 & \ldots & \ldots \\
 PSR J0251+2606        & J0251.0+2605  & 0.6 & \ldots & \ldots & \ldots & \ldots & \ldots & \ldots & \ldots & \ldots \\
 PSR J0312-0921        & J0312.1-0921  & 0.7 & \ldots & \ldots & \ldots & \ldots & \ldots & \ldots & \ldots & \ldots \\
 4FGL J0327.3+2355     & J0327.3+2355c & 5.8 & \ldots & \ldots & \ldots & \ldots & \ldots & \ldots & \ldots & \ldots \\
 4FGL J0336.0+7502     & J0336.0+7502  & 0.4 & \ldots & \ldots & \ldots & \ldots & J033610.2+750317 & 0.37 & \ldots & \ldots \\
 4FGL J0407.7-5702     & J0407.7-5702  & 2.4 & J040731.7-570024 & 0.91 & \ldots & \ldots & \ldots & \ldots & J040731.7-570026 & 0.7 \\
 3FGL J0427.9-6704     & J0427.8-6704  & 0.3 & J042749.6-670434 & 0.08 & 118029 & 0.2 & \ldots & \ldots & \ldots & \ldots \\
 1FGL J0523.5-2529     & J0523.3-2527  & 0.6 & \ldots & \ldots & 81522  & 1.8 & \ldots & \ldots & J052316.7-252738 & 3.3 \\
 4FGL J0540.0-7552     & J0540.0-7552  & 1.6 & J054001.7-755419 & 0.55 & 164275 & 3.2 & \ldots & \ldots & J054001.5-755414 & 4.9 \\
 PSR J0610-2100        & J0610.2-2100  & 0.4 & \ldots & \ldots & \ldots & \ldots & \ldots & \ldots & \ldots & \ldots \\
 PSR J0636+5128        & \ldots & \ldots & J063604.9+512900 & 0.73 & \ldots & \ldots & \ldots & \ldots & \ldots & \ldots \\
 4FGL J0639.1-8009     & J0639.1-8009  & 5.2 & J064059.5-801125 & 0.54 & 51549  & 1.5 & \ldots & \ldots & J064100.6-801127 & 3.3 \\
 4FGL J0705.8-0004     & J0705.8-0004  & 4.6 & \ldots & \ldots & \ldots & \ldots & \ldots & \ldots & \ldots &  \ldots     \\
 3FGL J0737.2-3233     & J0736.9-3231  & 1.1 & \ldots & \ldots & \ldots & \ldots & \ldots & \ldots & \ldots & \ldots \\
 PSR J0838-2827        & J0838.7-2827  & 1.1 & J083850.3-282756 & 1.01 & 90984  & 2.7 & \ldots & \ldots & \ldots & \ldots \\
 2FGL J0846.0+2820     & \ldots &     & \ldots & \ldots & \ldots & \ldots & J084621.8+280840 & 0.58 & \ldots & \ldots \\
 4FGL J0935.3+0901     & J0935.3+0901  & 1.3 & J093520.6+090035 & 0.37 & 171414 & 4.1 & \ldots & \ldots & J093520.8+090031 & 5.3 \\
 4FGL J0940.3-7610     & J0940.3-7610  & 0.8 & \ldots & \ldots & 178934 & 1.3 & \ldots & \ldots & J094023.1-761000 & 2.5 \\
 PSR J0952-0607        & J0952.1-0607  & 0.9 & J095208.3-060724 & 0.81 & \ldots & \ldots & \ldots & \ldots & \ldots & \ldots \\
 PSR J0955-3947        & J0955.3-3949  & 1.6 & \ldots & \ldots & 125466 & 0.5 & \ldots & \ldots & J095527.7-394752 & 1.1 \\
 PSR J1023+0038        & J1023.7+0038  & 0.4 & J102347.6+003841 & 0.18 & 6002   & 0.4 & J102347.6+003840 & 0.09 & J102347.6+003838 & 2.6 \\
 PSR J1036-4353        & J1036.6-4349  & 3.9 & \ldots & \ldots & \ldots & \ldots & \ldots & \ldots & \ldots & \ldots \\
 PSR J1048+2339        & J1048.6+2340  & 1.0 & \ldots & \ldots & 135807 & 1.2 & J104843.4+233953 & 0.2  & \ldots & \ldots \\
 CXOU J110926.4-650224 & J1110.3-6501  & 6.1 & J110926.4-650225 & 0.81 & 126443 & 2.4 & J110926.4-650225 & 0.03 & J110926.1-650224 & 1.6 \\
 PSR J1124-3653        & J1124.0-3653  & 0.8 & \ldots & \ldots & 139201 & 4.7 & J112401.1-365319 & 0.06 & \ldots & \ldots \\
 PSR J1221-0633        & J1221.4-0634  & 1.5 & \ldots & \ldots & \ldots & \ldots & \ldots & \ldots & \ldots & \ldots \\
 PSR J1227-4853        & J1228.0-4853  & 0.9 & J122758.7-485342 & 0.32 & 27586  & 1.9 & J122758.7-485343 & 0.26 & J122758.7-485342 & 0.7 \\
 PSR J1301+0833        & J1301.6+0834  & 0.5 & \ldots & \ldots  & 118543 & 4.1 & \ldots & \ldots & J130138.4+083402 & 5.9 \\
 PSR J1302-3258        & J1302.4-3258  & 1.0 & \ldots & \ldots & \ldots & \ldots & J130225.5-325837 & 0.21 & \ldots & \ldots \\
 PSR J1306-4035        & J1306.8-4035  & 1.2 & J130656.2-403523 & 0.26 & 51912  & 2.2 & \ldots & \ldots & J130656.1-403522 & 1.8 \\
 PSR J1311-3430        & J1311.7-3430  & 0.4 & J131145.7-343030 & 0.17 & 37289  & 1.8 & J131145.7-343030 & 0.38 & \ldots & \ldots \\
 PSR J1317-0157        & J1317.5-0153  & 4.5 & \ldots & \ldots &  \ldots &  \ldots & \ldots & \ldots & \ldots & \ldots \\
 PSR J1356+0230        & J1356.6+0234  & 4.2 & \ldots & \ldots &  \ldots &  \ldots & \ldots & \ldots & \ldots & \ldots \\
 4FGL J1408.6-2917     & J1408.6-2917  & 5.6 & J140826.8-292221 & 1.28 & \ldots & \ldots & \ldots & \ldots & \ldots & \ldots \\
 PSR J1417-4402        & J1417.6-4403  & 1.6 & \ldots & \ldots & 95132  & 1.9 & J141730.5-440257 & 0.41 & J141730.4-440257 & 1.8 \\
 PSR J1431-4715        & J1431.4-4711  & 4.6 & J143144.4-471524 & 3.41 & \ldots & \ldots & \ldots & \ldots &  \ldots & \ldots \\
 PSR J1446-4701        & J1446.6-4701  & 0.3 & J144635.8-470126 & 0.96 & \ldots & \ldots & \ldots & \ldots &  \ldots & \ldots \\
 PSR J1513-2550        & J1513.4-2549  & 1.6 & J151323.3-255029 & 1.36 & \ldots & \ldots & \ldots & \ldots &  \ldots & \ldots \\
 PSR J1544+4937        & J1544.0+4939  & 1.2 &  \ldots & \ldots  & \ldots & \ldots & \ldots & \ldots &  \ldots & \ldots \\
 4FGL J1544.2-2554     & J1544.2-2554  & 1.0 &  \ldots & \ldots  & \ldots & \ldots & \ldots & \ldots &  \ldots & \ldots \\
 3FGL J1544.6-1125     & J1544.5-1126  & 2.0 & J154439.4-112805 & 1.63 & 25260  & 2.1 & \ldots & \ldots & J154439.4-112802 & 2.1 \\
 PSR J1555-2908        & J1555.7-2908  & 0.4 & \ldots & \ldots & \ldots & \ldots & \ldots & \ldots & \ldots & \ldots \\
 PSR J1622-0315        & J1623.0-0315  & 0.3 & J162259.6-031538 & 0.86 & \ldots & \ldots & \ldots & \ldots & \ldots & \ldots \\
 PSR J1627+3219        & J1627.7+3219  & 1.8 & \ldots & \ldots & \ldots & \ldots & \ldots & \ldots & \ldots & \ldots \\
 PSR J1628-3205        & J1628.1-3204  & 1.4 & \ldots & \ldots & 110112 & 7.1 & J162806.9-320548 & 0.12 & \ldots & \ldots \\
 PSR J1641+8049        & J1641.2+8049  & 0.3 & \ldots & \ldots & \ldots & \ldots & \ldots & \ldots & \ldots & \ldots \\
 4FGL J1646.5-4406     & J1646.5-4406  & 1.4 & \ldots & \ldots & \ldots & \ldots & \ldots & \ldots & \ldots & \ldots \\
 PSR J1653-0158        & J1653.6-0158  & 0.5 & J165338.0-015837 & 0.56 & 97886  & 0.3 & J165338.0-015836 & 0.09 & \ldots & \ldots \\
 4FGL J1701.8-2226     & J1701.8-2226  & 5.0 &  \ldots & \ldots & \ldots & \ldots & \ldots &  \ldots &  \ldots & \ldots \\
 4FGL J1702.7-5655     & J1702.7-5655  & 0.8 &  \ldots & \ldots & \ldots & \ldots & \ldots &  \ldots &  \ldots & \ldots \\
 PSR J1705-1903        & J1705.6-1909  & 6.5 &  \ldots & \ldots & \ldots & \ldots & \ldots &  \ldots &  \ldots & \ldots \\
 PSR J1723-2837        & \ldots & \ldots & J172323.1-283757 & 0.48 & 56362  & 1.5 & J172323.1-283757 & 0.45 & J172323.1-283758 & 1.6 \\
 PSR J1731-1847        & J1731.7-1850  & 6.3 & J173117.6-184733 & 0.93 & \ldots & \ldots & J173117.5-184732 & 0.08 & \ldots & \ldots \\
 PSR J1745+1017        & J1745.5+1017  & 0.5 &  \ldots &  \ldots  & \ldots & \ldots & \ldots & \ldots & \ldots & \ldots \\
 PSR J1803-6707        & J1803.1-6708  & 0.7 &  \ldots &  \ldots  & \ldots & \ldots & \ldots & \ldots & J180304.4-670734 & 1.9 \\
 PSR J1805+0615        & J1805.6+0615  & 0.1 &  \ldots &  \ldots  & \ldots & \ldots & \ldots & \ldots & \ldots & \ldots \\
 PSR J1810+1744        & J1810.5+1744  & 0.4 &  \ldots &  \ldots  & \ldots & \ldots & J181037.3+174437 & 0.3  & \ldots & \ldots \\
 4FGL J1813.5+2819     & J1813.5+2819  & 1.8 &  \ldots &  \ldots  & \ldots & \ldots & \ldots & \ldots & \ldots & \ldots \\
 PSR J1816+4510        & J1816.5+4510  & 0.2 &  \ldots &  \ldots  & \ldots & \ldots & J181635.9+451033 & 0.34 & \ldots & \ldots \\
 4FGL J1819.4-1102     & J1819.4-1102  & 5.4 &  \ldots &  \ldots  & \ldots & \ldots & \ldots  & \ldots & \ldots & \ldots \\
 4FGL J1824.2+1231     & J1824.2+1231  & 1.8 &  \ldots &  \ldots  & \ldots & \ldots & \ldots  & \ldots & \ldots & \ldots \\
 PSR J1833-3840        & J1833.0-3840  & 0.7 &  \ldots &  \ldots  & \ldots & \ldots & \ldots  & \ldots & \ldots & \ldots \\
 4FGL J1838.2+3223     & J1838.2+3223  & 0.2 &  \ldots &  \ldots  & \ldots & \ldots & \ldots  & \ldots & \ldots & \ldots \\
 4FGL J1853.6-0620     & J1853.6-0620  & 5.0 &  \ldots &  \ldots  & \ldots & \ldots & \ldots  & \ldots & \ldots & \ldots \\
 4FGL J1859.2-0706     & J1859.2-0706  & 3.6 &  \ldots &  \ldots  & \ldots & \ldots & \ldots  & \ldots & \ldots & \ldots \\
 4FGL J1901.8-0718     & J1901.8-0718  & 2.2 &  \ldots &  \ldots  & \ldots & \ldots & \ldots  & \ldots & \ldots & \ldots \\
 PSR J1908+2105        & J1908.9+2103  & 1.4 & J190857.5+210502 & 3.0  & \ldots & \ldots & \ldots & \ldots & \ldots & \ldots \\
 PSR J1910-5320        & J1910.7-5320  & 0.7 & \ldots & \ldots & \ldots & \ldots & J191049.1-532057 & 0.46 & \ldots & \ldots \\
 PSR J1946-5403        & J1946.5-5402  & 0.8 & J194634.4-540343 & 0.55 & \ldots & \ldots & \ldots & \ldots & \ldots & \ldots \\
 PSR J1947-1120        & J1947.6-1121  & 1.5 & \ldots &  \ldots & \ldots & \ldots & \ldots & \ldots & \ldots & \ldots \\
 PSR J1957+2516        & J1957.3+2517  & 3.7 & \ldots &  \ldots & \ldots & \ldots & \ldots & \ldots & \ldots & \ldots \\
 PSR B1957+20          & J1959.5+2048  & 0.3 & J195936.7+204815 & 0.1  & \ldots & \ldots & J195936.7+204814 & 0.53 & \ldots & \ldots \\
 PSR J2017-1614        & J2017.7-1612  & 2.0 & J201746.0-161416 & 1.02 & \ldots & \ldots & \ldots & \ldots & \ldots & \ldots \\
 PSR J2039-5617        & J2039.5-5617  & 0.1 & J203935.0-561709 & 0.74 & 31211  & 1.8 & \ldots & \ldots & \ldots & \ldots \\
 PSR J2047+1053        & J2047.3+1051  & 3.2 & \ldots & \ldots & \ldots & \ldots & J204710.2+105307 & 0.12 & \ldots & \ldots \\
 PSR J2051-0827        & J2051.0-0826  & 1.8 & J205107.4-082737 & 0.31 & \ldots & \ldots & J205107.5-082737 & 0.18 & \ldots & \ldots \\
 PSR J2052+1219        & J2052.7+1218  & 1.1 &  \ldots & \ldots & \ldots & \ldots & \ldots & \ldots & \ldots & \ldots \\
 4FGL J2054.2+6904     & J2054.2+6904  & 1.7 &  \ldots & \ldots & 87966  & 1.3 & \ldots & \ldots & \ldots & \ldots \\
 PSR J2055+1545        & J2055.8+1545  & 0.4 &  \ldots & \ldots & \ldots & \ldots & \ldots & \ldots & \ldots & \ldots \\
 PSR J2115+5448        & J2115.1+5449  & 0.7 & J211511.7+544844 & 0.27 & \ldots & \ldots & \ldots & \ldots & \ldots & \ldots \\
 PSR J2129-0429        & J2129.8-0428  & 1.6 & J212945.0-042906 & 0.16 & 82190  & 0.6 & J212945.0-042906 & 0.17 & \ldots & \ldots \\
 PSR J2214+3000        & J2214.6+3000  & 0.5 & J221438.8+300038 & 0.19 & \ldots & \ldots & J221438.8+300038 & 0.08 & \ldots & \ldots \\
 PSR J2215+5135        & J2215.6+5135  & 0.7 & J221532.6+513536 & 0.41 & 75117  & 4.5 & J221532.7+513536 & 0.16 & \ldots & \ldots \\
 PSR J2234+0944        & J2234.7+0943  & 0.6 & \ldots & \ldots & \ldots & \ldots & \ldots & \ldots & \ldots & \ldots \\
 PSR J2241-5236        & J2241.7-5236  & 0.5 & J224142.0-523635 & 0.42 & 167739 & 3.0 & J224142.0-523636 & 0.46 & \ldots & \ldots \\
 PSR J2256-1024        & J2256.8-1024  & 1.1 & \ldots & \ldots & \ldots & \ldots & J225656.3-102434 & 0.34 & \ldots & \ldots \\
 PSR J2333-5526        & J2333.1-5527  & 1.8 & J233316.0-552620 & 0.45 & 143919 & 6.0 & \ldots & \ldots & \ldots & \ldots \\
 PSR J2339-0533        & J2339.6-0533  & 0.1 & J233938.7-053305 & 0.27 & 54830  & 1.1 & J233938.7-053305 & 0.2  & \ldots & \ldots \\
\hline
\enddata
\tablenotetext{}{{\bf Notes.} See Section~\ref{sec:xg} for details. (This table is available in machine-readable form in the online article.)}
\end{deluxetable*}

\begin{figure*}
\plotone{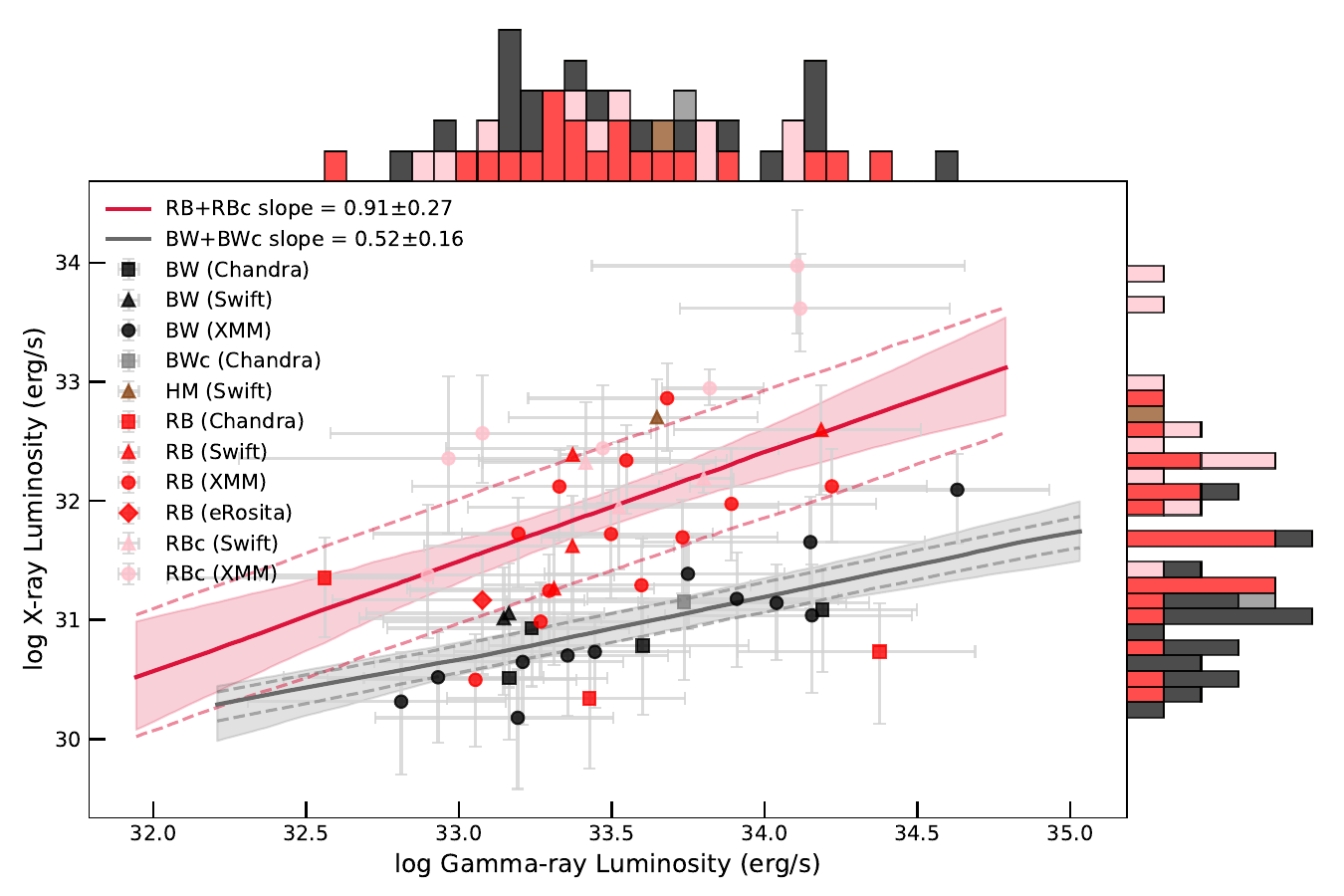}
\caption{The relationship between X-ray ($L_{\rm X}$) and $\gamma$-ray ($L_{\gamma}$) luminosities for spider systems. $\gamma$-ray luminosities are not significantly different between spider types. On the other hand, RBs exhibit systematically higher X-ray luminosities likely due to stronger IBSs. BWs follow a relation of $L_{\rm X} \propto L_{\gamma}^{0.5\pm0.2}$, while RBs may follow a steeper trend of $L_{\rm X} \propto L_{\gamma}^{0.9\pm0.3}$, suggesting differing emission contributions across populations.
\label{fig:x_gamma_lum}}
\end{figure*}

\begin{figure}
\plotone{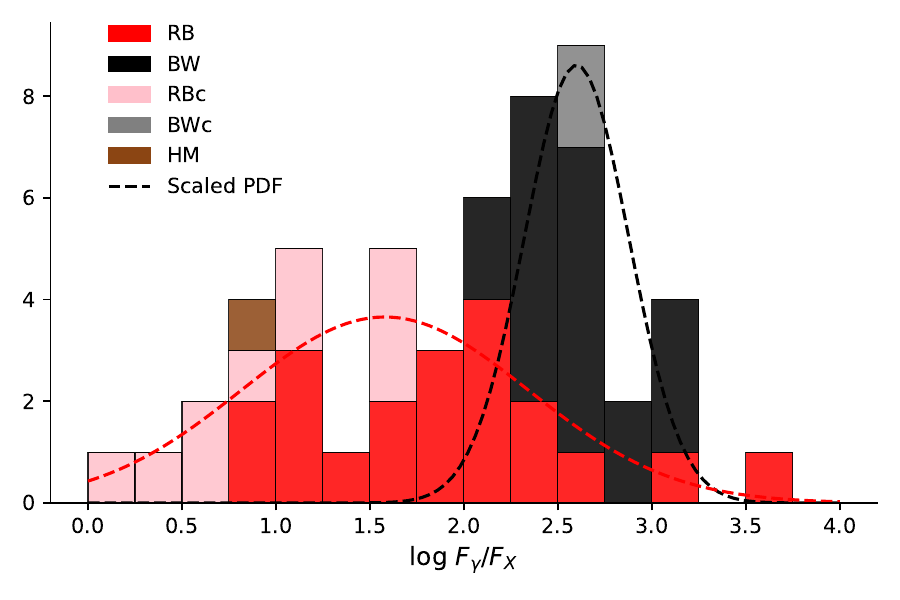}
\caption{Histogram of the logarithm of the ratio of $\gamma$-ray ($F_{\gamma}$) to X-ray ($F_{\rm X}$) fluxes for spider systems (bin size: 0.25 dex), color coded by spider type. The flux ratios for BWs are more narrowly distributed and generally higher. For $\log \, (F_{\gamma}/F_{\rm X}) < 2$, it becomes increasingly likely that the system is an RB.
\label{fig:fg_fx}}
\end{figure}

For the X-ray catalogs, we performed the counterpart search using search radii approximately corresponding to the 2$\sigma$ average positional uncertainties reported for each catalog. Specifically, we used a radius of 4$\arcsec$.4 for 4XMM-DR14, 2$\arcsec$ for CSC2.1, 8$\arcsec$ for 2SXPS, and 9$\arcsec$.2 for eRASS1. All identified counterparts are unique within their respective search radii, and no sources have multiple matches.

We associate a total of 86 Fermi 4FGL sources and 55 X-ray counterparts with spider systems (see Table~\ref{tab:recap}), corresponding to 77\% and 50\% of the catalog, respectively. Some sources appear in earlier Fermi catalogs (2FGL~J0846.0$+$2820, \citealt{ssj+17}; and possibly PSR~J1723$-$2837, \citealt{hui14}), or in Fermi 3PC catalog (PSR~J0630$+$5128, \citealt{saa+23}) but are absent from the 4FGL catalog, suggesting intermittent and/or weak $\gamma$-ray emission. Including these three sources raises the $\gamma$-ray association rate among spiders to 80\%. 

Twenty-two sources (20\%) lack high-energy counterparts. Of these, 12 originate from the FAST-GPPS survey \citep{why+24} and suffer from low-precision radio positions and high Galactic background. Three are BWs with very large DM-inferred distances (PSR J2055$+$3829, PSR J1928$+$1245, and PSR J1745$-$23), and one is an RB, PSR J1932$+$2121, which has a relatively large Gaia parallax-based distance estimate ($\sim$5~kpc). These characteristics likely place them beyond the sensitivity limits of current surveys. PSR J1720$-$0533, also a BW without high-energy detection, is probably located farther away ($\sim$3~kpc) than suggested by its DM-derived distance \citep{koljonen24}. Two sources are spider candidates based on optical variability (3FGL J2221.6$+$6507 and ZTF J1406$+$1222). This leaves three sources with moderate distances ($\lesssim$2.5~kpc), two BWs (PSR J1602$-$1009 and PSR J1630$+$3550) and one RB (PSR J1242$-$4712), which represent promising targets for deeper X-ray and $\gamma$-ray follow-up observations aimed at identifying potential high-energy counterparts.

We computed X-ray and $\gamma$-ray luminosities using the X-ray fluxes from the aforementioned catalogs and the distances described in Section~\ref{sec:distances}. The $\gamma$-ray energy fluxes are quoted in the Fermi-LAT 0.1--100~GeV energy band, obtained by spectral fitting (as tabulated in 4FGL-DR4). For X-ray luminosity estimates, fluxes were used in the following priority order: XMM-Newton (0.2--12~keV), Neil Gehrels Swift Observatory (0.3--10~keV), Chandra X-ray Observatory (0.2--7keV), and eROSITA onboard Spektrum-Roentgen-Gamma (0.2--8~keV). The resulting luminosities and the observatory used in each case are provided in Table~\ref{tab:dist_lum}.

Figure~\ref{fig:x_gamma_lum} displays the X-ray and $\gamma$-ray luminosities of the spider systems. The $\gamma$-ray luminosities span $10^{32}$–$10^{35}$ erg s$^{-1}$ with a mean luminosity between $10^{33}$ and $10^{33.5}$ erg s$^{-1}$. No significant differences are observed between spider subtypes in their $\gamma$-ray luminosities.
%  and follow an approximately log-normal distribution (Shapiro-Wilk $W = 0.98$, $P<0.21$)

Spider X-ray luminosities, known to exhibit a bimodal distribution \citep{lee18, ssc+22, koljonen23}, range from $10^{30}$ to $10^{34}$ erg s$^{-1}$. RBs consistently display higher X-ray luminosities, with values exceeding $10^{32}$ erg s$^{-1}$ found exclusively among them (see Figure~\ref{fig:x_gamma_lum}). This enhanced emission likely originates from stronger IBSs \citep[e.g.,][]{romani16, wadiasingh17, wadiasingh18, vandermerwe20}. Additionally, in RBs, the IBS tends to wrap around the pulsar rather than the companion  \citep{wadiasingh17, kandel19, sullivan25}. This geometry places the shock closer to the pulsar, resulting in a higher incident particle flux and more efficient conversion of spin-down power into X-rays, further contributing to the observed luminosity differences between the two populations.

Both BWs (and candidates) and RBs (and candidates) exhibit correlations between their X-ray and $\gamma$-ray luminosities (Figure~\ref{fig:x_gamma_lum}). This correlation is expected if the spin-down power is the primary driver of both emissions. In this framework, $\gamma$-rays are thought to originate close to the light cylinder, while X-rays arise from synchrotron emission of relativistic $e^\pm$ pairs accelerated at the termination shock of the pulsar wind. The relationship between spin-down power and X-ray or $\gamma$-ray luminosity in spiders has been previously studied, e.g., in \citet{lee18,hui19,koljonen23}. BWs follow the relation $L_{\rm X} \propto L_{\gamma}^{0.6\pm0.2}$, while RBs may follow a steeper relation of $L_{\rm X} \propto L_{\gamma}^{1.0\pm0.3}$. The potential difference in slope (though consistent within 1$\sigma$) suggests that X-ray emission in RBs becomes relatively more efficient than in BWs at higher $\gamma$-ray luminosities. 

In Figure~\ref{fig:fg_fx}, we present a histogram of the logarithm of the $\gamma$-ray to X-ray flux (or luminosity) ratios. BWs exhibit a much narrower distribution, with a mean value of $\log \, (F_{\gamma}/F_{\rm X}) = 2.6 \pm 0.3$, whereas RBs show a broader distribution, centered at $\log \, (F_{\gamma}/F_{\rm X}) = 1.6 \pm 0.8$. The lowest observed ratio for BWs is $\log \, (F_{\gamma}/F_{\rm X}) \approx 2.1$; sources with lower values are thus likely to be RBs. Therefore, the flux ratio can serve as a diagnostic tool for distinguishing between source types when both X-ray and $\gamma$-ray fluxes are available. This represents an update to the flux ratios reported by \citet{salvetti17}, in which such distinctions could not be clearly identified. 
 
\subsubsection{Optical and Infrared} \label{sec:oir}

\startlongtable
\begin{deluxetable*}{lllllllll}  
\tabletypesize{\scriptsize}
\tablewidth{0pt}
\tablecaption{Optical/infrared counterparts. See Section~\ref{sec:oir} for details. \label{tab:oir_counterparts}}
\tablehead{\colhead{Name} & \colhead{Gaia ID} & \colhead{sep} & \colhead{Pan-STARRS ID} & \colhead{sep} & \colhead{SkyMapper ID} & \colhead{sep} & \colhead{2MASS ID} & \colhead{sep} \\ \colhead{} & \colhead{(Gaia DR3)} & \colhead{('')} & \colhead{(PS1)} & \colhead{('')} & \colhead{(SMSS)} & \colhead{('')} & \colhead{(2MASS)} & \colhead{('')}}
\startdata
\hline
 PSR J0212+5321        & 455282205716288384  & 0.02          & 172030330436573611 & 0.02          & \ldots             & \ldots        & 02121047+5321387 & 0.10          \\
 4FGL J0327.3+2355     & 68712810947091072   & 0.37          & 136630519292105092 & 0.37          & \ldots             & \ldots        & \ldots           & \ldots        \\
 4FGL J0336.0+7502     & 544927450310303104  & 0.00          & 198060540424576487 & 0.47          & \ldots             & \ldots        & \ldots           & \ldots        \\
 4FGL J0407.7-5702     & 4682464743003293312 & 0.03          & \ldots             & \ldots        & 040731.71-570025.3 & 0.20          & \ldots           & \ldots        \\
 3FGL J0427.9-6704     & 4656677385699742208 & 0.22          & \ldots             & \ldots        & 042749.62-670435.2 & 0.69          & \ldots           & \ldots        \\
 1FGL J0523.5-2529     & 2957031626919939456 & 0.23          & 77440808205337977  & 0.22          & 052316.92-252737.1 & 0.47          & 05231692-2527369 & 0.09          \\
 4FGL J0540.0-7552     & 4648562676357022208 & 0.01          & \ldots             & \ldots        & 054001.87-755419.4 & 0.18          & \ldots           & \ldots        \\
 4FGL J0639.1-8009     & 5207836863615934080 & 0.00          & \ldots             & \ldots        & \ldots             & \ldots        & \ldots           & \ldots        \\
 4FGL J0705.8-0004     & 3112746864531854336 & 0.00          & 108001064703258353 & 0.00          & 070552.88+000023.6 & 0.10          & \ldots           & \ldots        \\
 3FGL J0737.2-3233     & 5592027220171664128 & 0.26          & \ldots             & \ldots        & 073656.22-323255.0 & 0.36          & 07365621-3232551 & 0.16          \\
 PSR J0838-2827        & 5645504747023158400 & 0.02          & 73841297100561268  & 0.03          & \ldots             & \ldots        & \ldots           & \ldots        \\
 2FGL J0846.0+2820     & 705098703608575744  & 0.29          & 141771315911514198 & 0.30          & \ldots             & \ldots        & 08462187+2808408 & 0.23          \\
 4FGL J0935.3+0901     & 588191888537402112  & 0.00          & 118811438363042433 & 0.03          & \ldots             & \ldots        & \ldots           & \ldots        \\
 4FGL J0940.3-7610     & 5203822684102798592 & 0.02          & \ldots             & \ldots        & 094023.63-761001.0 & \textbf{1.06} & 09402362-7610012 & \textbf{1.26} \\
 PSR J0952-0607        & \ldots              & \ldots        & 100651480346382461 & 0.09          & \ldots             & \ldots        & \ldots           & \ldots        \\
 PSR J0955-3947        & 5419965878188457984 & 0.02          & \ldots             & \ldots        & 095527.82-394752.2 & 0.06          & \ldots           & \ldots        \\
 PSR J1023+0038        & 3831382647922429952 & 0.12          & 108771559487124073 & 0.10          & 102347.70+003840.8 & 0.14          & 10234768+0038412 & 0.43          \\
 PSR J1036-4353        & 5367876720979404288 & 0.05          & \ldots             & \ldots        & 103630.23-435308.7 & 0.11          & \ldots           & \ldots        \\
 PSR J1048+2339        & 3990037124929068032 & 0.06          & 136391621809818375 & 0.04          & \ldots             & \ldots        & \ldots           & \ldots        \\
 CXOU J110926.4-650224 & 5240167590731178624 & 0.23          & \ldots             & \ldots        & \ldots             & \ldots        & \ldots           & \ldots        \\
 PSR J1227-4853        & 6128369984328414336 & 0.04          & \ldots             & \ldots        & 122758.74-485342.6 & 0.11          & 12275874-4853428 & 0.31          \\
 PSR J1301+0833        & \ldots              & \ldots        & 118271954094819640 & 0.07          & \ldots             & \ldots        & \ldots           & \ldots        \\
 PSR J1306-4035        & 6140785016794586752 & 0.33          & \ldots             & \ldots        & 130656.29-403523.4 & 0.36          & 13065627-4035233 & 0.28          \\
 PSR J1311-3430        & 6179115508262195200 & 0.03          & \ldots             & \ldots        & 131145.72-343030.2 & 0.07          & \ldots           & \ldots        \\
 ZTF J1406+1222        & 1226507286664241280 & 0.06          & 122852117342624990 & 0.71          & \ldots             & \ldots        & \ldots           & \ldots        \\
 4FGL J1408.6-2917     & 6173249201411182592 & 0.01          & 72752121116613280  & 0.05          & 140826.84-292221.1 & \textbf{0.91} & \ldots           & \ldots        \\
 PSR J1417-4402        & 6096705840454620800 & 0.41          & \ldots             & \ldots        & 141730.57-440257.5 & 0.37          & 14173057-4402574 & 0.33          \\
 PSR J1431-4715        & 6098156298150016768 & 0.10          & \ldots             & \ldots        & 143144.62-471527.5 & 0.08          & \ldots           & \ldots        \\
 4FGL J1544.2-2554     & 6235002859670996352 & 0.07          & 76882360643979456  & 0.09          & \ldots             & \ldots        & 0.1              & \ldots        \\
 3FGL J1544.6-1125     & 6268529198286308224 & 0.59          & 94232361640998794  & 0.58          & 154439.39-112804.8 & 0.76          & \ldots           & \ldots        \\
 PSR J1555-2908        & 6041127310076589056 & 0.02          & 73032389194490840  & 0.20          & \ldots             & \ldots        & \ldots           & \ldots        \\
 PSR J1622-0315        & 4358428942492430336 & 0.06          & 104082457484918020 & 0.06          & 162259.63-031537.2 & 0.11          & \ldots           & \ldots        \\
 PSR J1628-3205        & 6025344817107454464 & 0.14          & 69482470291653997  & 0.13          & 162807.01-320548.9 & 0.16          & \ldots           & \ldots        \\
 PSR J1653-0158        & 4379227476242700928 & 0.11          & 105622534086488129 & 0.09          & \ldots             & \ldots        & \ldots           & \ldots        \\
 4FGL J1701.8-2226     & 4113999849726115072 & 0.17          & 80972554479621675  & 0.17          & 170147.50-223125.9 & 0.31          & \ldots           & \ldots        \\
 PSR J1705-1903        & 4128524776292614912 & \textbf{0.99} & 85122564325326863  & \textbf{1.01} & \ldots             & \ldots        & \ldots           & \ldots        \\
 PSR J1723-2837        & 4059795674516044800 & 0.45          & 73642608465751153  & 0.40          & 172323.17-283757.7 & 0.53          & 17232318-2837571 & 0.03          \\
 PSR J1731-1847        & 4121864828231575168 & \textbf{1.66} & 85442628229009307  & \textbf{1.67} & \ldots             & \ldots        & \ldots           & \ldots        \\
 PSR J1745-23          & 4068800644840738816 & \textbf{1.05} & 79902663754330387  & \textbf{1.34} & \ldots             & \ldots        & \ldots           & \ldots        \\
 PSR J1803-6707        & 6436867623955512064 & 0.03          & \ldots             & \ldots        & 180304.23-670735.9 & 0.20          & \ldots           & \ldots        \\
 PSR J1805+0615        & \ldots              & \ldots        & 115502714266536702 & 0.10          & \ldots             & \ldots        & \ldots           & \ldots        \\
 PSR J1810+1744        & 4526229058440076288 & 0.11          & 129292726553492988 & 0.11          & \ldots             & \ldots        & \ldots           & \ldots        \\
 4FGL J1813.5+2819     & 4589333153195186432 & 0.06          & 142002733594883208 & 0.05          & \ldots             & \ldots        & \ldots           & \ldots        \\
 PSR J1816+4510        & 2115337192179377792 & 0.05          & 162212741497611947 & 0.05          & \ldots             & \ldots        & \ldots           & \ldots        \\
 4FGL J1819.4-1102     & 4154315810078443392 & 0.02          & 94722749521506396  & 0.03          & \ldots             & \ldots        & 18194851-1103418 & 0.06          \\
 4FGL J1824.2+1231     & 4485064648767923328 & 0.06          & 123052760370011669 & 0.08          & \ldots             & \ldots        & \ldots           & \ldots        \\
 PSR J1830-0106g       & \ldots              & \ldots        & 106682775293960515 & \textbf{0.91} & \ldots             & \ldots        & \ldots           & \ldots        \\
 4FGL J1838.2+3223     & 2090923983890463104 & 0.02          & 146882795700574433 & 0.13          & \ldots             & \ldots        & \ldots           & \ldots        \\
 4FGL J1853.6-0620     & 4253939878633735296 & 0.03          & 100422833485776802 & 0.03          & \ldots             & \ldots        & \ldots           & \ldots        \\
 4FGL J1859.2-0706     & 4205576146028340864 & 0.04          & 99402848273340334  & 0.10          & \ldots             & \ldots        & \ldots           & \ldots        \\
 4FGL J1901.8-0718     & 4205374935376962304 & 0.06          & 99262854837553437  & 0.06          & 190156.20-071652.3 & \textbf{1.77} & \ldots           & \ldots        \\
 PSR J1908+2105        & 4519819661567533696 & 0.62          & 133302872386881340 & 0.61          & \ldots             & \ldots        & \ldots           & \ldots        \\
 PSR J1910-5320        & 6644467032871428992 & 0.00          & \ldots             & \ldots        & 191049.12-532057.2 & 0.13          & \ldots           & \ldots        \\
 PSR J1919+1502g       & 4320684258808495360 & \textbf{2.13} & \ldots             & \ldots        & \ldots             & \ldots        & \ldots           & \ldots        \\
 PSR J1928+1245        & 4316237348443952128 & 0.13          & 123312921891428291 & 0.13          & 192845.40+124553.2 & 0.27          & 19284537+1245536 & 0.36          \\
 PSR J1932+2121        & 2018081407303262208 & \textbf{1.07} & 133622930887172807 & \textbf{1.08} & \ldots             & \ldots        & \ldots           & \ldots        \\
 PSR J1947-1120        & 4189956032809439488 & 0.03          & 94392969093081360  & 0.03          & 194738.23-112027.1 & 0.08          & 19473823-1120271 & 0.07          \\
 PSR J1957+2516        & 1834595731470345472 & 0.03          & 138322993939420960 & \textbf{1.31} & \ldots             & \ldots        & \ldots           & \ldots        \\
 PSR B1957+20          & 1823773960079217024 & 0.75          & 132962999031405508 & 0.52          & \ldots             & \ldots        & \ldots           & \ldots        \\
 PSR J2017-1614        & \ldots              & \ldots        & 88513044422425287  & 0.05          & \ldots             & \ldots        & \ldots           & \ldots        \\
 PSR J2039-5617        & 6469722508861870080 & 0.01          & \ldots             & \ldots        & 203934.98-561709.2 & 0.10          & \ldots           & \ldots        \\
 PSR J2047+1053        & \ldots              & \ldots        & 121063117926563305 & 0.16          & \ldots             & \ldots        & \ldots           & \ldots        \\
 PSR J2052+1219        & \ldots              & \ldots        & 122803131991300216 & 0.18          & \ldots             & \ldots        & \ldots           & \ldots        \\
 4FGL J2054.2+6904     & 2271107409667918080 & 0.02          & 190903134956837351 & 0.03          & \ldots             & \ldots        & \ldots           & \ldots        \\
 PSR J2055+3829        & 1872588462410154240 & \textbf{1.61} & 154193137931590438 & \textbf{1.58} & \ldots             & \ldots        & \ldots           & \ldots        \\
 PSR J2055+1545        & 1763537692275731328 & 0.01          & 126903139493667392 & 0.02          & \ldots             & \ldots        & \ldots           & \ldots        \\
 PSR J2129-0429        & 2672030065446134656 & 0.15          & 102613224377138157 & 0.13          & 212945.06-042906.7 & 0.13          & 21294505-0429070 & 0.19          \\
 PSR J2214+3000        & \ldots              & \ldots        & 144013336619243286 & 0.09          & \ldots             & \ldots        & \ldots           & \ldots        \\
 PSR J2215+5135        & 2001168543319218048 & 0.03          & 169913338861152867 & 0.02          & \ldots             & \ldots        & \ldots           & \ldots        \\
 3FGL J2221.6+6507     & 2206544469143391872 & 0.31          & 186003356364777687 & 0.33          & \ldots             & \ldots        & 22223274+6500207 & 0.41          \\
 PSR J2256-1024        & \ldots              & \ldots        & 95503442349308910  & 0.02          & \ldots             & \ldots        & \ldots           & \ldots        \\
 PSR J2333-5526        & 6496325574947304448 & 0.02          & \ldots             & \ldots        & 233315.95-552620.8 & 0.19          & \ldots           & \ldots        \\
 PSR J2339-0533        & 2440660623886405504 & 0.18          & 101333549113918682 & 0.17          & 233938.74-053305.2 & 0.20          & \ldots           & \ldots        \\
\hline
\enddata
\tablenotetext{}{{\bf Notes.} See Section~\ref{sec:oir} for details. Bold values in the Sep. columns indicate angular separations greater than 0$\arcsec$.9. In these cases, the counterpart association may be incorrect. (This table is available in machine-readable form in the online article.)}
\end{deluxetable*}

\begin{figure*}
\plotone{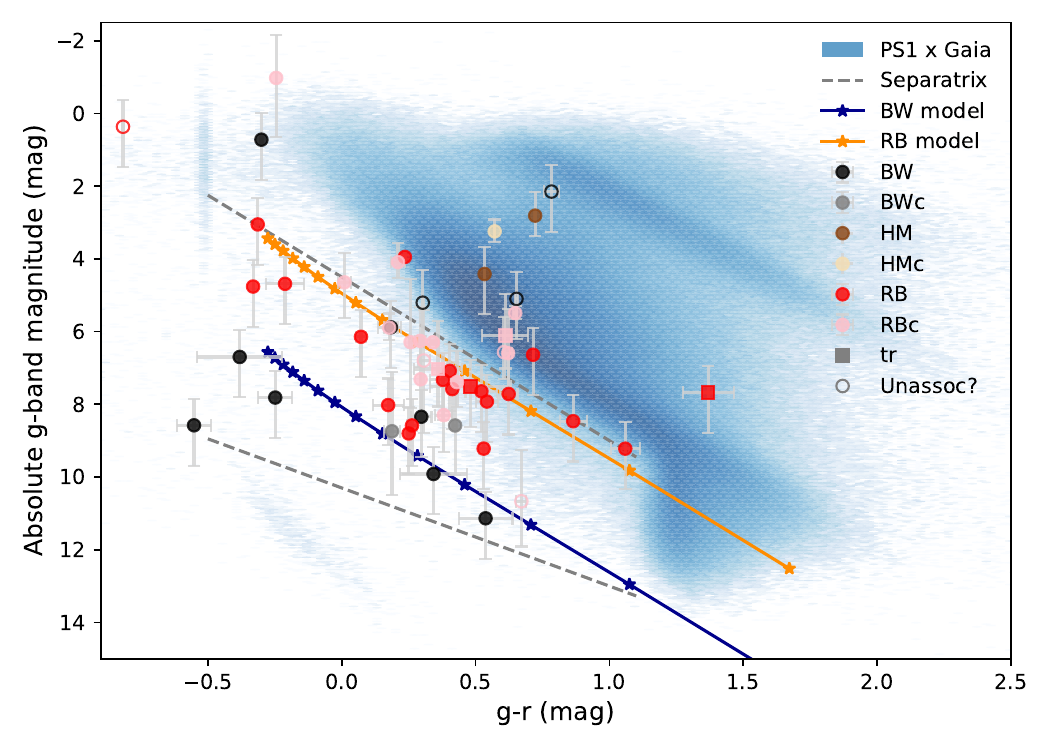}
\caption{Color-magnitude diagram showing extinction-corrected absolute $g$-band magnitudes vs. $g–r$ color for spider counterparts, overlaid on a background of three million Pan-STARRS field stars crossmatched with Gaia to obtain accurate distance and absolute magnitude measurements. Spider systems predominantly occupy a region between the main sequence and WD branch. Outliers are, in most cases, either HM systems (with giant companions), tMSPs (which may include optical emission from an accretion disk), or potentially unassociated stars (where the optical counterpart is uncertain due to a large angular separation from the pulsar or candidate). Most spider systems are located between the two dashed lines, where the upper boundary is defined by $M_g = 4.5 \times (g - r) + 4.5$, and the lower by $M_g = 2.7 \times (g - r) + 10.3$. We also draw two representative blackbody tracks with different temperatures from 3000 K to 16,000 K separated by 1000 K (marked as stars) for BW-type (with radius $R_{\rm BW}=0.09 \, R_{\odot}$) and RB-type stars (with radius $R_{\rm RB}=0.38 \, R_{\odot}$).
\label{fig:color_mag}}
\end{figure*}

We identify optical counterparts to 67 spider systems, representing 61\% of the sample (see Table~\ref{tab:recap}). These include 63 Gaia, 54 Pan-STARRS, and 29 SkyMapper counterparts. As expected, the majority (72\%) correspond to RBs, which host larger and brighter companion stars. Additionally, we find 15 2MASS counterparts.

Unlike high-energy counterparts, the reliability of optical associations is more uncertain, particularly in regions of high stellar density. Therefore, we caution that not all associations may be genuine. Although we used a 2$\arcsec$ search radius to accommodate high proper motions or positional uncertainties, separations exceeding $\sim$1$\arcsec$ are questionable unless corroborated by other evidence (e.g., orbital period matching across wave bands). To highlight this, we mark in bold the angular separations larger than 0$\arcsec$.9 in Table~\ref{tab:oir_counterparts}.

The magnitudes reported from Pan-STARRS and SkyMapper are mean values from catalog photometry. Pan-STARRS typically provides 12 epochs per filter over a 5 yr survey, with single-epoch exposure times of 30--45 s, while SkyMapper typically provides six epochs per filter over a 7 yr survey, with single-epoch exposures of 100 s. Taken at random orbital phases, these observations generally average over the orbit and thus include flux contributions from both the intrinsic companion and the heated face. However, for faint sources, the detections may preferentially sample orbital phases near maximum brightness. We calculated the apparent and absolute optical magnitudes using Pan-STARRS and SkyMapper data, along with distances from Section~\ref{sec:distances}. We corrected the optical magnitudes for extinction using reddening estimates from \citet{schlegel98} and extinction coefficients from \citet{schlafly11} for Pan-STARRS, as well as those provided for SkyMapper on their website.\footnote{\url{https://skymapper.anu.edu.au/filter-transformations/}} RB counterparts have extinction-corrected median $g$-, $r$-, $i$-, and $z$-band magnitudes of 18.5, 18.3, 18.1, and 17.8, with interquartile ranges of 17.3--19.4, 16.5--19.0, 16.2--18.6, and 16.0--18.5, respectively. BWs are on average $\sim$2 mag fainter, with medians of 20.2, 19.9, 20.2, and 19.5 and interquartile ranges of 19.1--20.8, 19.0--21.1, 19.1--21.2, and 18.4--20.3, respectively. The median $g–r$ color is 0.38 for RBs (interquartile range: 0.18--0.61) and 0.07 for BWs (interquartile range: $-$0.32--0.31). The individual values are reported in Table~\ref{tab:dist_lum}.

Figure~\ref{fig:color_mag} shows a color–magnitude diagram of spider systems, overlaid on a representative background of three million randomly selected Gaia stars with well-determined distances and magnitudes,\footnote{Selection based on DR3, following the criteria in \citet[][Appendix B]{gaia18}.} crossmatched with the Pan-STARRS catalog to provide consistent magnitudes and colors. Spider systems generally occupy a region between the main sequence and the WD branch. Similar to what was done with Gaia data by \citet{antoniadis21}, we empirically define the region where the majority of spiders (75\%) are found by drawing two lines: the upper boundary is given by $M_g = 4.5 \times (g - r) + 4.5$, and the lower by $M_g = 2.7 \times (g - r) + 10.3$. These lines serve as a rough guide for identifying potential companion candidates. However, for colors $(g - r) < 0.25$, this linear approximation becomes less accurate, as the main sequence curves upward and spider systems appear above the defined region. Most outliers (10 out of 17 systems) are either HM systems with giant companions, tMSPs with possible optical emission from an accretion disk, or systems where the optical counterpart is uncertain due to a large angular separation from the pulsar or candidate coordinates. We note that the putative optical counterpart to the BW PSR~J1928$+$1245 has an anomalously high $g$-band absolute magnitude of $M_g = 0.7^{+1.1}_{-0.7}$ mag. This is the highest among confirmed spider systems and is approximately 6 mag brighter than BWs at a similar color. Such discrepancy strongly suggests that this source is not the true counterpart, but rather an unrelated object along the line of sight to the pulsar, as already suggested by \citet{koljonen23}.   

We also compute two representative color–magnitude tracks for BW- and RB-type companion stars, assuming blackbody spectra with temperatures ranging from 3000 K to 16,000 K. These are shown in Figure~\ref{fig:color_mag} as blue and orange lines, with star symbols marking blackbody temperatures in steps of 1000 K; cooler stars correspond to higher $(g - r)$ values. We adopt stellar radii by assuming the volume-equivalent Roche-lobe radius \citep{eggleton83}, a neutron star mass of $1.8 \, M_{\odot}$, mass ratios $q_{\rm BW} = 0.01$ and $q_{\rm RB} = 0.3$, median orbital periods of $P_{\rm b,BW} = 3.6$ hr and $P_{\rm b,RB} = 6.2$ hr (see Section~\ref{sec:orbits}), and a volume-equivalent Roche-lobe filling factor of $f = 0.6$. This results in $R_{\rm BW} = 0.09 \, R_{\odot}$ and $R_{\rm RB} = 0.38 \, R_{\odot}$ for BWs and RBs, respectively. Synthetic magnitudes are calculated using Pan-STARRS filter curves, assuming no extinction. These tracks agree quite well with the location of the majority of spider systems.    

\startlongtable
\begin{deluxetable*}{lcccccccccc}
\tabletypesize{\scriptsize}
\tablewidth{0pt}
\tablecaption{Spider Distances and Luminosities \label{tab:dist_lum}}
\tablehead{\colhead{Name} & \colhead{d} & \colhead{Method\tablenotemark{a}} & \colhead{$F_{\rm X}$} & \colhead{Det.\tablenotemark{b}} & \colhead{$L_{\rm X}$} & \colhead{$F_{\rm \gamma}$} & \colhead{$L_{\rm \gamma}$} & \colhead{$g$} & \colhead{$M_{g}$} & \colhead{$g-r$} \\ \colhead{} & \colhead{(kpc)} & \colhead{} & \colhead{($10^{-14}$ } & \colhead{} & \colhead{($10^{31}$ } & \colhead{($10^{-12}$ } & \colhead{($10^{33}$ } & \colhead{(mag)} & \colhead{(mag)} & \colhead{(mag)} \\ \colhead{} & \colhead{} & \colhead{} & \colhead{erg/s/cm$^{2}$)} & \colhead{} & \colhead{erg/s)} & \colhead{erg/s/cm$^{2}$)} & \colhead{erg/s)} & \colhead{} & \colhead{} & \colhead{}}
\startdata
\hline
 PSR J0023+0923        & $\sim$1.2              & YMW16  & $1.7\pm0.3$    & Chandra & $0.3^{+0.4}_{-0.2}$        & $7.8\pm0.6$  & $1.5^{+1.6}_{-1.0}$    & \ldots & \ldots & \ldots \\
 PSR J0212+5321        & $1.12^{+0.03}_{-0.03}$ & Gaia   & $162.7\pm18.4$ & Swift   & $24.5^{+4.2}_{-3.8}$       & $15.6\pm0.6$ & $2.4^{+0.2}_{-0.2}$    & 14.20 & $3.95^{+0.05}_{-0.06}$ & 0.24  \\
 PSR J0251+2606        & $\sim$1.2              & YMW16  & \ldots & \ldots & \ldots & $4.9\pm0.4$  & $0.8^{+0.9}_{-0.5}$    & \ldots & \ldots & \ldots \\
 PSR J0312-0921        & $\sim$0.8              & YMW16  & \ldots & \ldots & \ldots & $5.5\pm0.4$  & $0.4^{+0.5}_{-0.3}$    & \ldots & \ldots & \ldots \\
 4FGL J0327.3+2355     & $1.8^{+0.9}_{-0.5}$    & Gaia   & \ldots & \ldots & \ldots & $2.2\pm0.6$  & $0.9^{+1.5}_{-0.5}$    & 18.67 & $7.4^{+0.7}_{-0.9}$    & 0.43  \\
 4FGL J0336.0+7502     & $2.4^{+1.9}_{-1.1}$    & Gaia   & $2.1\pm0.4$    & Chandra & $1.4^{+4.1}_{-1.1}$        & $8.1\pm0.5$  & $5.5^{+13.3}_{-4.0}$   & 20.46 & $8.6^{+1.4}_{-1.3}$    & 0.42  \\
 4FGL J0407.7-5702     & $2.5^{+1.8}_{-0.9}$    & Gaia   & $49.6\pm1.3$   & XMM     & $37.4^{+75.7}_{-23.2}$     & $1.6\pm0.3$  & $1.2^{+3.0}_{-0.8}$    & 20.29 & $8.3^{+1.0}_{-1.2}$    & 0.38  \\
 3FGL J0427.9-6704     & $2.5^{+0.5}_{-0.4}$    & Gaia   & $117.6\pm1.9$  & XMM     & $88.4^{+40.1}_{-24.4}$     & $8.8\pm0.5$  & $6.6^{+3.3}_{-2.0}$    & 17.89 & $5.9^{+0.3}_{-0.4}$    & 0.18  \\
 1FGL J0523.5-2529     & $2.1^{+0.2}_{-0.2}$    & Gaia   & $29.2\pm3.0$   & Swift   & $15.4^{+4.9}_{-3.7}$       & $12.0\pm0.5$ & $6.3^{+1.6}_{-1.2}$    & 17.10 & $5.5^{+0.2}_{-0.2}$    & 0.65  \\
 4FGL J0540.0-7552     & $2.6^{+2.1}_{-1.1}$    & Gaia   & $33.8\pm1.0$   & XMM     & $27.8^{+65.7}_{-18.1}$     & $3.6\pm0.5$  & $3.0^{+8.1}_{-2.1}$    & \ldots & \ldots & \ldots \\
 PSR J0541+2959g       & $\sim$1.5              & YMW16  & \ldots & \ldots & \ldots & \ldots & \ldots & \ldots & \ldots & \ldots \\
 PSR J0610-2100        & $\sim$3.3              & YMW16  & \ldots & \ldots & \ldots & $7.2\pm0.5$  & $9.2^{+10.0}_{-6.1}$   & \ldots & \ldots & \ldots \\
 PSR J0636+5128        & $\sim$0.5              & NE2001 & $2.8\pm0.4$    & XMM     & $0.08^{+0.10}_{-0.06}$     & \ldots & \ldots & \ldots & \ldots & \ldots \\
 4FGL J0639.1-8009     & $1.8^{+1.9}_{-0.9}$    & Gaia   & $60.6\pm6.1$   & XMM     & $23.0^{+86.7}_{-17.7}$     & $2.5\pm0.5$  & $0.9^{+3.9}_{-0.7}$    & \ldots & \ldots & \ldots \\
 4FGL J0705.8-0004     & $2.4^{+1.8}_{-0.9}$    & Gaia   & \ldots & \ldots & \ldots & $3.2\pm0.8$  & $2.1^{+6.0}_{-1.5}$    & 18.46 & $6.6^{+1.0}_{-1.2}$    & 0.62  \\
 3FGL J0737.2-3233     & $0.66^{+0.02}_{-0.02}$ & Gaia   & \ldots & \ldots & \ldots & $10.2\pm1.1$ & $0.52^{+0.10}_{-0.08}$ & 15.38 & $6.29^{+0.07}_{-0.07}$ & 0.34  \\
 PSR J0838-2827        & $3.0^{+2.0}_{-1.2}$    & Gaia   & $8.9\pm0.6$    & XMM     & $9.4^{+18.6}_{-6.2}$       & $7.3\pm0.5$  & $7.7^{+15.5}_{-5.1}$   & 20.08 & $7.7^{+1.1}_{-1.1}$    & 0.62  \\
 2FGL J0846.0+2820     & $3.7^{+0.6}_{-0.5}$    & Gaia   & $1.1\pm0.4$    & Chandra & $1.8^{+1.4}_{-0.9}$        & \ldots & \ldots & 16.11 & $3.2^{+0.3}_{-0.3}$    & 0.57  \\
 4FGL J0935.3+0901     & $1.2^{+1.1}_{-0.5}$    & Gaia   & $13.8\pm1.1$   & XMM     & $2.4^{+7.1}_{-1.7}$        & $4.6\pm0.5$  & $0.8^{+2.5}_{-0.6}$    & 21.07 & $10.7^{+1.2}_{-1.4}$   & 0.67  \\
 4FGL J0940.3-7610     & $1.9^{+1.0}_{-0.5}$    & Gaia   & $50.4\pm15.7$  & Swift   & $21.0^{+46.2}_{-13.6}$     & $6.2\pm0.6$  & $2.6^{+4.3}_{-1.4}$    & 17.92 & $6.6^{+0.7}_{-1.0}$    & 0.61  \\
 PSR J0952-0607        & $\sim$1.7              & YMW16  & $0.9\pm0.2$    & XMM     & $0.3^{+0.5}_{-0.2}$        & $2.4\pm0.3$  & $0.9^{+1.0}_{-0.6}$    & \ldots & \ldots & \ldots \\
 PSR J0955-3947        & $\sim$3.3              & NE2001 & $29.9\pm6.1$   & Swift   & $39.6^{+53.8}_{-28.2}$     & $11.6\pm0.9$ & $15.3^{+17.1}_{-10.2}$ & 17.30 & $4.7^{+1.1}_{-0.7}$    & -0.21 \\
 PSR J1023+0038        & $1.4^{+0.1}_{-0.1}$    & Gaia   & $492.1\pm1.2$  & XMM     & $114.4^{+26.3}_{-19.6}$    & $32.4\pm0.8$ & $7.5^{+1.9}_{-1.4}$    & 17.91 & $7.2^{+0.2}_{-0.2}$    & 1.37  \\
 PSR J1036-4353        & $\sim$2.1              & NE2001 & \ldots & \ldots & \ldots & $3.4\pm0.5$  & $1.8^{+2.3}_{-1.3}$    & 19.63 & $8.0^{+1.1}_{-0.7}$    & 0.17  \\
 PSR J1048+2339        & $\sim$2.0              & YMW16  & $8.7\pm2.9$    & Swift   & $4.2^{+6.8}_{-3.2}$        & $4.9\pm0.5$  & $2.3^{+2.7}_{-1.6}$    & 19.97 & $8.5^{+1.1}_{-0.7}$    & 0.86  \\
 CXOU J110926.4-650224 & $4.7^{+3.3}_{-2.2}$    & Gaia   & $355.6\pm2.7$  & XMM     & $939.1^{+1806.5}_{-684.8}$ & $4.8\pm1.0$  & $12.7^{+32.1}_{-10.0}$ & \ldots & \ldots   & \ldots \\
 PSR J1124-3653        & $\sim$1.0              & YMW16  & $9.8\pm3.3$    & Swift   & $1.1^{+1.8}_{-0.9}$        & $12.5\pm0.6$ & $1.5^{+1.5}_{-1.0}$    & \ldots & \ldots   & \ldots \\
 PSR J1221-0633        & $\sim$1.3              & YMW16  & \ldots & \ldots & \ldots & $5.8\pm0.5$  & $1.1^{+1.2}_{-0.7}$    & \ldots & \ldots   & \ldots \\
 PSR J1227-4853        & $\sim$1.2              & YMW16  & $118.3\pm1.1$  & XMM     & $21.9^{+21.4}_{-14.1}$     & $19.0\pm1.6$ & $3.5^{+4.0}_{-2.4}$    & 17.98 & $7.5^{+1.1}_{-0.7}$    & 0.48  \\
 PSR J1242-4712        & $\sim$2.5              & YMW16  & \ldots & \ldots & \ldots & \ldots & \ldots & \ldots & \ldots   & \ldots \\
 PSR J1301+0833        & $\sim$1.2              & YMW16  & $5.7\pm2.1$    & Swift   & $1.0^{+1.7}_{-0.8}$        & $7.7\pm0.5$  & $1.4^{+1.5}_{-0.9}$    & \ldots & \ldots   & \ldots \\
 PSR J1302-3258        & $\sim$1.4              & YMW16  & $0.9\pm0.3$    & Chandra & $0.2^{+0.3}_{-0.2}$        & $10.9\pm0.5$ & $2.7^{+2.8}_{-1.8}$    & \ldots & \ldots   & \ldots \\
 PSR J1306-4035        & $\sim$1.4              & YMW16  & $55.6\pm1.5$   & XMM     & $13.2^{+13.4}_{-8.6}$      & $9.0\pm0.7$  & $2.1^{+2.4}_{-1.4}$    & 18.39 & $7.6^{+1.1}_{-0.7}$    & 0.52  \\
 PSR J1311-3430        & $\sim$2.4              & YMW16  & $17.6\pm0.4$   & XMM     & $12.4^{+12.5}_{-8.0}$      & $60.6\pm1.2$ & $42.6^{+42.6}_{-27.6}$ & 18.62 & $6.7^{+1.1}_{-0.7}$    & -0.38 \\
 PSR J1317-0157        & $\sim$2.8              & NE2001 & \ldots & \ldots & \ldots & $1.5\pm0.3$  & $1.4^{+1.9}_{-1.0}$    & \ldots & \ldots & \ldots \\
 PSR J1356+0230        & $\sim$1.8              & YMW16  & \ldots & \ldots & \ldots & $1.9\pm0.3$  & $0.8^{+1.0}_{-0.5}$    & \ldots & \ldots & \ldots \\
 ZTF J1406+1222        &  \ldots & \ldots & \ldots & \ldots & \ldots & \ldots & \ldots & 19.83 & \ldots & -0.06 \\
 4FGL J1408.6-2917     &  \ldots & \ldots & $1.3\pm0.8$    & XMM     & \ldots & $5.2\pm0.7$  & \ldots & 20.96 & \ldots & -0.10 \\
 PSR J1417-4402        & $\sim$2.2              & YMW16  & $90.0\pm6.0$   & Swift   & $50.4^{+55.0}_{-33.5}$     & $7.9\pm0.7$  & $4.4^{+5.0}_{-3.0}$    & 16.09 & $4.4^{+1.1}_{-0.7}$    & 0.53  \\
 PSR J1431-4715        & $\sim$1.8              & YMW16  & $2.4\pm0.2$    & XMM     & $1.0^{+1.1}_{-0.6}$        & $4.7\pm0.7$  & $1.8^{+2.3}_{-1.3}$    & 17.45 & $6.1^{+1.1}_{-0.7}$    & 0.07  \\
 PSR J1446-4701        & $\sim$1.6              & YMW16  & $1.7\pm0.2$    & XMM     & $0.5^{+0.6}_{-0.3}$        & $7.7\pm0.7$  & $2.3^{+2.6}_{-1.5}$    & \ldots & \ldots   & \ldots \\
 PSR J1513-2550        & $\sim$4.0              & YMW16  & $0.6\pm0.2$    & XMM     & $1.1^{+1.9}_{-0.8}$        & $7.6\pm0.6$  & $14.3^{+16.0}_{-9.6}$  & \ldots & \ldots   & \ldots \\
 PSR J1544+4937        & $\sim$3.0              & YMW16  & \ldots & \ldots & \ldots & $2.4\pm0.3$  & $2.5^{+3.0}_{-1.7}$    & \ldots & \ldots   & \ldots \\
 4FGL J1544.2-2554     & \ldots & \ldots & \ldots & \ldots & \ldots & $7.6\pm0.7$  & \ldots & 20.18 &  \ldots  & 0.19  \\
 3FGL J1544.6-1125     & $3.1^{+2.1}_{-1.1}$    & Gaia   & $359.9\pm2.3$  & XMM     & $414.9^{+774.1}_{-235.6}$  & $11.4\pm0.9$ & $13.1^{+27.0}_{-7.8}$  & 18.57 & $6.1^{+0.9}_{-1.1}$    & 0.61  \\
 PSR J1555-2908        & $\sim$7.6              & YMW16  & \ldots &  \ldots & \ldots & $4.7\pm0.6$  & $31.9^{+38.7}_{-21.9}$ & 20.27 & $5.9^{+1.1}_{-0.7}$    & 0.18  \\
 PSR J1602-1009        & \ldots & \ldots & \ldots &  \ldots & \ldots & \ldots & \ldots & \ldots & \ldots & \ldots \\
 PSR J1622-0315        & $\sim$1.1              & YMW16  & $2.0\pm0.5$    & XMM     & $0.3^{+0.4}_{-0.2}$        & $7.2\pm0.7$  & $1.1^{+1.3}_{-0.8}$    & 18.87 & $8.6^{+1.1}_{-0.7}$    & 0.26  \\
 PSR J1627+3219        & \ldots & \ldots & \ldots & \ldots & \ldots & $3.6\pm0.3$  & \ldots & \ldots & \ldots & \ldots \\
 PSR J1628-3205        & $\sim$1.2              & YMW16  & $10.2\pm3.7$   & Swift   & $1.8^{+3.1}_{-1.4}$        & $11.3\pm0.9$ & $2.0^{+2.3}_{-1.4}$    & \ldots & \ldots   & \ldots \\
 PSR J1630+3550        & $\sim$1.6              & YMW16  & \ldots & \ldots & \ldots & \ldots & \ldots & \ldots & \ldots   & \ldots \\
 PSR J1641+8049        & $\sim$3.0              & YMW16  & \ldots & \ldots & \ldots & $2.0\pm0.3$  & $2.2^{+2.8}_{-1.5}$    & \ldots & \ldots   & \ldots \\
 4FGL J1646.5-4406     & \ldots & \ldots & \ldots & \ldots & \ldots & $8.5\pm3.4$  & \ldots  \ldots & \ldots & \ldots   & \ldots \\
 PSR J1653-0158        & $1.6^{+2.9}_{-1.0}$    & Gaia   & $17.4\pm0.8$   & XMM     & $5.7^{+39.9}_{-4.7}$       & $34.3\pm1.0$ & $11.1^{+77.4}_{-9.3}$  & 20.16 & $9.1^{+1.9}_{-2.2}$    & 0.17  \\
 4FGL J1701.8-2226     & $2.9^{+3.5}_{-1.6}$    & Gaia   & \ldots & \ldots &  \ldots & $2.6\pm0.7$  & $2.6^{+13.5}_{-2.2}$   & 18.57 & $6.3^{+1.8}_{-1.7}$    & 0.26  \\
 4FGL J1702.7-5655     & \ldots & \ldots & \ldots & \ldots &  \ldots & $28.8\pm1.4$ & \ldots & \ldots & \ldots & \ldots \\
 PSR J1705-1903        & $\sim$2.3              & YMW16  & \ldots & \ldots &  \ldots & $3.6\pm0.8$  & $2.4^{+3.3}_{-1.7}$    & 19.02 & $7.2^{+1.1}_{-0.7}$    & 0.30  \\
 PSR J1720-0533        & $\sim$1.3              & NE2001 & \ldots & \ldots &  \ldots & \ldots & \ldots & \ldots & \ldots & \ldots \\
 PSR J1723-2837        & $0.91^{+0.04}_{-0.03}$ & Gaia   & $229.3\pm2.8$  & XMM     & $22.5^{+2.2}_{-2.0}$       & \ldots & \ldots & 12.34 & $2.55^{+0.09}_{-0.09}$ & -0.31 \\
 PSR J1731-1847        & $\sim$4.8              & YMW16  & $1.6\pm0.4$    & XMM     & $4.5^{+6.4}_{-3.3}$        & $5.2\pm1.1$  & $14.1^{+19.3}_{-10.1}$ & 18.50 & $5.1^{+1.1}_{-0.7}$    & 0.65  \\
 PSR J1745-23          & $\sim$7.9              & YMW16  & \ldots & \ldots & \ldots & \ldots & \ldots & 16.65 & $2.1^{+1.1}_{-0.7}$    & 0.78  \\
 PSR J1745+1017        & $\sim$1.2              & YMW16  & \ldots & \ldots & \ldots & $7.6\pm0.6$  & $1.3^{+1.5}_{-0.9}$    & \ldots & \ldots & \ldots \\
 PSR J1803-6707        & $\sim$1.4              & YMW16  & $5.9\pm2.0$    & eRosita & $1.5^{+2.4}_{-1.1}$        & $4.8\pm0.5$  & $1.2^{+1.4}_{-0.8}$    & 20.02 & $9.2^{+1.1}_{-0.7}$    & 0.53  \\
 PSR J1805+0615        & $\sim$3.9              & YMW16  & \ldots & \ldots & \ldots & $5.3\pm0.5$  & $9.5^{+11.1}_{-6.4}$   & -0.76 & \ldots & \ldots \\
 PSR J1810+1744        & $\sim$2.4              & YMW16  & $1.8\pm0.3$    & Chandra & $1.2^{+1.6}_{-0.8}$        & $23.2\pm0.9$ & $15.5^{+16.1}_{-10.1}$ & 19.68 & $7.8^{+1.1}_{-0.7}$    & -0.25 \\
 4FGL J1813.5+2819     & $3.2^{+1.7}_{-0.9}$    & Gaia   & \ldots & \ldots &  \ldots & $2.3\pm0.3$  & $2.9^{+4.9}_{-1.6}$    & 18.79 & $6.3^{+0.7}_{-0.9}$    & 0.29  \\
 PSR J1814+0045g       & $\sim$8.9              & YMW16  & \ldots & \ldots &  \ldots & \ldots & \ldots & \ldots & \ldots & \ldots \\
 PSR J1816+4510        & $\sim$4.3              & YMW16  & $0.2\pm0.1$    & Chandra & $0.5^{+0.8}_{-0.4}$        & $10.6\pm0.5$ & $23.8^{+25.0}_{-15.6}$ & 17.95 & $4.8^{+1.1}_{-0.7}$    & -0.33 \\
 4FGL J1819.4-1102     & $4.5^{+3.3}_{-2.4}$    & Gaia   & \ldots & \ldots & \ldots & $8.2\pm2.8$  & $19.7^{+59.2}_{-16.8}$ & 12.28 & $-1.0^{+1.6}_{-1.2}$   & -0.24 \\
 4FGL J1824.2+1231     & $2.4^{+2.0}_{-0.9}$    & Gaia   & \ldots & \ldots & \ldots & $3.3\pm0.9$  & $2.2^{+7.2}_{-1.6}$    & 18.90 & $7.0^{+1.0}_{-1.3}$    & 0.36  \\
 PSR J1830-0106g       & $\sim$5.0              & YMW16  & \ldots & \ldots & \ldots & \ldots & \ldots & -5.87 & \ldots & \ldots \\
 PSR J1833-3840        & $\sim$4.7              & YMW16  & \ldots & \ldots & \ldots & $2.8\pm0.4$  & $7.3^{+9.3}_{-5.1}$    & \ldots & \ldots & \ldots \\
 4FGL J1838.2+3223     & $2.2^{+2.5}_{-1.2}$    & Gaia   & \ldots & \ldots & \ldots & $2.7\pm0.6$  & $1.5^{+6.9}_{-1.3}$    & 20.45 & $8.7^{+1.8}_{-1.6}$    & 0.19  \\
 PSR J1838+1507g       & $\sim$2.4              & YMW16  & \ldots & \ldots & \ldots & \ldots &  \ldots & \ldots & \ldots & \ldots \\
 PSR J1847+0342g       & $\sim$2.9              & YMW16  & \ldots & \ldots & \ldots & \ldots &  \ldots & \ldots & \ldots & \ldots \\
 PSR J1849+0304g       & $\sim$4.5              & YMW16  & \ldots & \ldots & \ldots & \ldots &  \ldots & \ldots & \ldots & \ldots \\
 4FGL J1853.6-0620     & $5.4^{+2.8}_{-2.3}$    & Gaia   & \ldots & \ldots & \ldots & $4.9\pm1.4$  & $17.0^{+33.7}_{-13.0}$ & -1.97 & \ldots & \ldots \\
 4FGL J1859.2-0706     & $5.8^{+2.6}_{-2.1}$    & Gaia   & \ldots & \ldots & \ldots & $4.8\pm1.1$  & $19.4^{+31.4}_{-13.4}$ & 18.46 & $4.6^{+1.0}_{-0.8}$    & 0.01  \\
 PSR J1859+0313g       & $\sim$3.1              & YMW16  & \ldots & \ldots & \ldots & \ldots & \ldots & \ldots & \ldots & \ldots \\
 4FGL J1901.8-0718     & $3.4^{+3.1}_{-2.1}$    & Gaia   & \ldots & \ldots & \ldots & $3.7\pm0.7$  & $5.2^{+17.0}_{-4.5}$   & 19.48 & $6.8^{+2.0}_{-1.4}$    & 0.31  \\
 PSR J1908+2105        & $\sim$2.6              & YMW16  & $2.4\pm0.6$    & XMM     & $2.0^{+2.9}_{-1.4}$        & $4.9\pm0.8$  & $4.0^{+5.0}_{-2.8}$    & 19.40 & $7.3^{+1.1}_{-0.7}$    & 0.38  \\
 PSR J1910-5320        & $\sim$1.0              & YMW16  & $19.1\pm2.1$   & Chandra & $2.3^{+2.7}_{-1.5}$        & $3.1\pm0.5$  & $0.4^{+0.5}_{-0.3}$    & 18.79 & $8.8^{+1.1}_{-0.7}$    & 0.25  \\
 PSR J1919+0126g       & $\sim$6.3              & YMW16  & \ldots & \ldots & \ldots & \ldots & \ldots & \ldots & \ldots & \ldots \\
 PSR J1919+1502g       & $\sim$6.0              & YMW16  & \ldots & \ldots & \ldots & \ldots & \ldots & \ldots & \ldots & \ldots \\
 PSR J1928+1245        & $\sim$6.1              & YMW16  & \ldots & \ldots & \ldots & \ldots & \ldots & 14.64 & $0.7^{+1.1}_{-0.7}$    & -0.30 \\
 PSR J1931+1428g       & $\sim$6.2              & YMW16  & \ldots & \ldots & \ldots & \ldots & \ldots & \ldots &  \ldots & \ldots \\
 PSR J1932+2121        & $\sim$5.1              & YMW16  & \ldots & \ldots & \ldots & \ldots & \ldots & 8.70  &  \ldots & -2.36 \\
 PSR J1946-5403        & $\sim$1.1              & YMW16  & $1.0\pm0.3$    & XMM     & $0.2^{+0.2}_{-0.1}$        & $9.8\pm0.5$  & $1.6^{+1.6}_{-1.0}$    & \ldots & \ldots & \ldots \\
 PSR J1947-1120        & \ldots & \ldots & \ldots & \ldots & \ldots & $3.3\pm0.6$  & \ldots & 16.59 & \ldots & 0.72  \\
 PSR J1953+1006g       & $\sim$2.5              & YMW16  & \ldots & \ldots & \ldots & \ldots & \ldots & \ldots & \ldots & \ldots \\
 PSR J1957+2516        & $\sim$2.7              & YMW16  & \ldots & \ldots & \ldots & $4.0\pm1.3$  & $3.4^{+5.5}_{-2.6}$    & 12.49 & $0.4^{+1.1}_{-0.7}$    & -0.82 \\
 PSR B1957+20          & $\sim$1.7              & YMW16  & $6.8\pm0.3$    & XMM     & $2.4^{+2.5}_{-1.6}$        & $15.7\pm0.9$ & $5.6^{+6.0}_{-3.7}$    & 19.53 & $8.3^{+1.1}_{-0.7}$    & 0.30  \\
 PSR J2003+3032g       & $\sim$6.8              & YMW16  & \ldots & \ldots & \ldots & \ldots & \ldots & \ldots & \ldots & \ldots \\
 PSR J2017-1614        & $\sim$1.4              & YMW16  & $1.8\pm0.3$    & XMM     & $0.4^{+0.6}_{-0.3}$        & $6.5\pm0.6$  & $1.6^{+1.8}_{-1.1}$    & -0.23 &                        & \ldots \\
 PSR J2039-5617        & $\sim$1.7              & YMW16  & $14.2\pm0.8$   & XMM     & $4.9^{+5.3}_{-3.3}$        & $15.4\pm0.6$ & $5.4^{+5.6}_{-3.5}$    & 18.74 & $7.6^{+1.1}_{-0.7}$    & 0.41  \\
 PSR J2047+1053        & $\sim$2.8              & YMW16  & $0.7\pm0.2$    & Chandra & $0.6^{+0.9}_{-0.4}$        & $4.3\pm0.6$  & $4.0^{+4.9}_{-2.7}$    & 22.15 & $9.9^{+1.1}_{-0.7}$    & 0.34  \\
 PSR J2051-0827        & $\sim$1.5              & YMW16  & $0.8\pm0.3$    & XMM     & $0.2^{+0.3}_{-0.2}$        & $2.5\pm0.3$  & $0.6^{+0.8}_{-0.4}$    & \ldots &                        & \ldots \\
 PSR J2052+1219        & $\sim$3.9              & YMW16  & \ldots & \ldots & \ldots & $4.6\pm0.6$  & $8.4^{+10.1}_{-5.7}$   & 21.54 & $8.6^{+1.1}_{-0.7}$    & -0.55 \\
 4FGL J2054.2+6904     & $2.6^{+1.9}_{-1.0}$    & Gaia   & $11.1\pm2.2$   & Swift   & $8.9^{+22.7}_{-6.3}$       & $4.2\pm0.5$  & $3.3^{+7.8}_{-2.3}$    & 19.38 & $7.3^{+1.1}_{-1.2}$    & 0.29  \\
 PSR J2055+3829        & $\sim$4.6              & YMW16  & \ldots & \ldots & \ldots & \ldots & \ldots & -2.21 & \ldots & \ldots \\
 PSR J2055+1545        & $\sim$3.7              & YMW16  & \ldots & \ldots & \ldots & $2.3\pm0.6$  & $3.7^{+5.4}_{-2.7}$    & 20.74 & $7.9^{+1.1}_{-0.7}$    & 0.54  \\
 PSR J2115+5448        & $\sim$3.1              & YMW16  & $1.3\pm0.3$    & XMM     & $1.5^{+2.2}_{-1.1}$        & $7.0\pm0.7$  & $8.1^{+9.3}_{-5.5}$    & \ldots & \ldots & \ldots \\
 PSR J2129-0429        & $2.0^{+0.3}_{-0.2}$    & Gaia   & $23.2\pm0.6$   & XMM     & $10.8^{+4.0}_{-2.6}$       & $6.8\pm0.5$  & $3.2^{+1.4}_{-0.9}$    & 17.34 & $5.9^{+0.3}_{-0.3}$    & 0.71  \\
 PSR J2214+3000        & $\sim$1.7              & YMW16  & $4.2\pm0.3$    & XMM     & $1.4^{+1.6}_{-0.9}$        & $32.6\pm0.7$ & $10.9^{+11.0}_{-7.1}$  & \ldots & \ldots & \ldots \\
 PSR J2215+5135        & $\sim$2.8              & YMW16  & $14.4\pm0.8$   & XMM     & $13.3^{+14.2}_{-8.7}$      & $18.0\pm0.8$ & $16.6^{+17.4}_{-10.9}$ & 19.30 & $7.1^{+1.1}_{-0.7}$    & 0.40  \\
 3FGL J2221.6+6507     & $2.3^{+0.6}_{-0.4}$    & Gaia   & \ldots & \ldots & \ldots & \ldots & \ldots & 15.85 & $4.1^{+0.4}_{-0.5}$    & 0.21  \\
 PSR J2234+0944        & $\sim$1.6              & YMW16  & \ldots & \ldots & \ldots & $10.0\pm0.6$ & $3.0^{+3.3}_{-2.0}$    & \ldots & \ldots & \ldots \\
 PSR J2241-5236        & $\sim$1.0              & YMW16  & $4.9\pm0.3$    & XMM     & $0.5^{+0.6}_{-0.4}$        & $25.0\pm1.1$ & $2.8^{+2.9}_{-1.8}$    & \ldots & \ldots & \ldots \\
 PSR J2256-1024        & $\sim$1.3              & YMW16  & $4.0\pm0.4$    & Chandra & $0.9^{+1.0}_{-0.6}$        & $8.2\pm0.5$  & $1.7^{+1.9}_{-1.1}$    & 21.75 & $11.1^{+1.1}_{-0.7}$   & 0.54  \\
 PSR J2333-5526        & $\sim$2.5              & YMW16  & $7.2\pm1.4$    & XMM     & $5.3^{+7.0}_{-3.7}$        & $4.3\pm0.4$  & $3.1^{+3.6}_{-2.1}$    & 21.18 & $9.2^{+1.1}_{-0.7}$    & 1.06  \\
 PSR J2339-0533        & $\sim$0.8              & YMW16  & $26.0\pm0.8$   & XMM     & $1.8^{+1.8}_{-1.1}$        & $29.2\pm0.8$ & $2.0^{+2.0}_{-1.3}$    & 18.15 & $8.8^{+1.1}_{-0.7}$    & -1.30 \\
\hline
\enddata
\tablenotetext{}{{\bf Notes.} (This table is available in machine-readable form in the online article.)}
\tablenotetext{a}{We estimate distances to spiders using dispersion measures (DMs) and the Galactic electron density model of \citet[][``YMW16'']{yao17}, or, in some cases, the older model of \citet[][``NE2001'']{cordes02}. When a DM is not available, we use geometric distances derived from Gaia Data Release 3 (DR3) parallaxes. However, if the relative parallax uncertainty is less than 0.2, we always adopt the parallax-based distance. The method used is noted in this column. See Section~\ref{sec:distances} for details.} 
\tablenotetext{b}{X-ray detectors and energy bands used to obtain the X-ray fluxes are as follows: XMM---EPIC on board XMM-Newton (0.2--12 keV); Swift---the X-ray Telescope (XRT) on board the Neil Gehrels Swift Observatory (0.3--10 keV); Chandra---ACIS on board the Chandra X-ray Observatory (0.2--7 keV); eRosita---eROSITA on board Spektrum-Roentgen-Gamma (0.2--8 keV).}
\end{deluxetable*}

\subsection{Spin-orbit} \label{sec:spin-orbit}

\begin{figure*}
\gridline{\fig{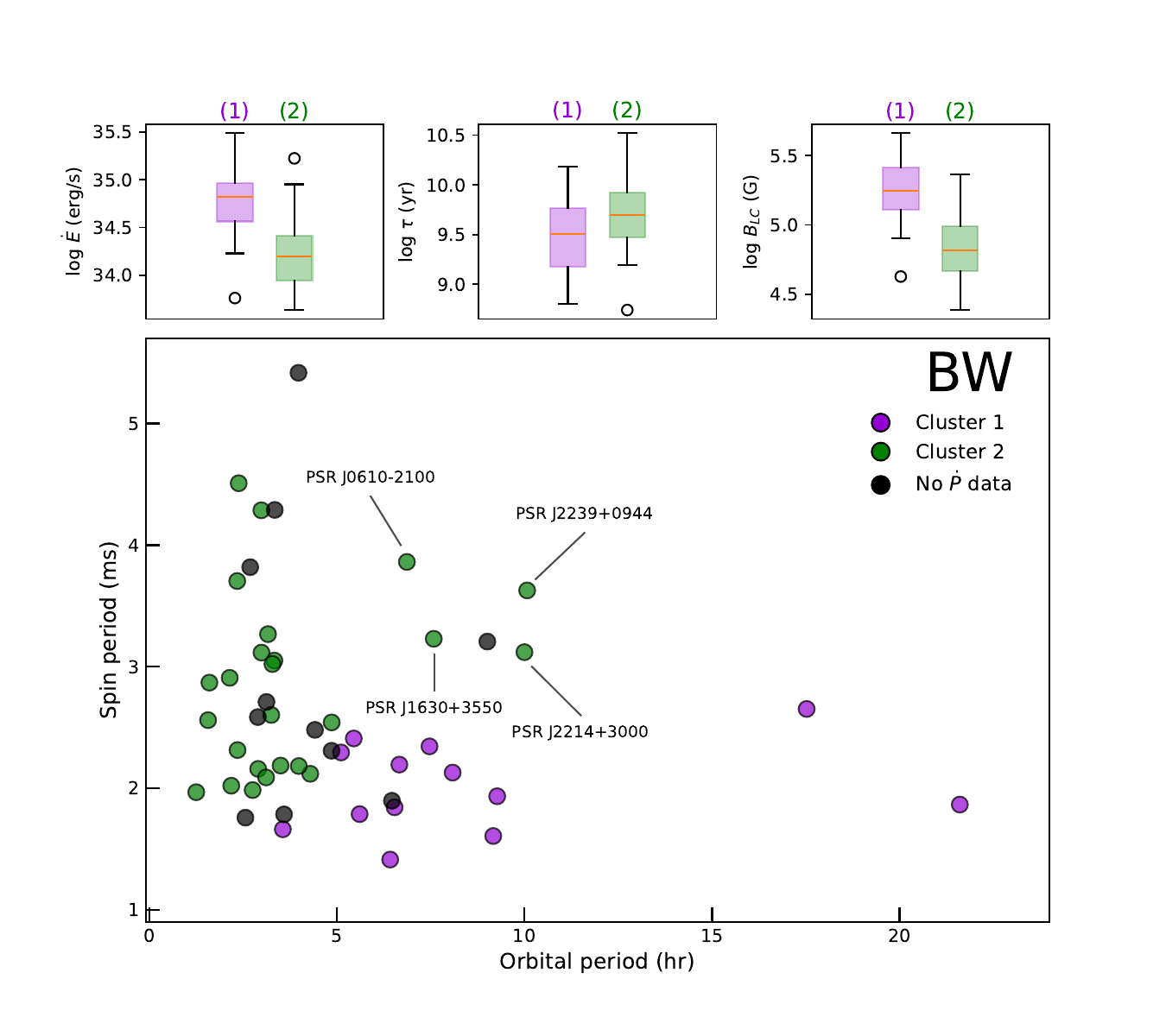}{0.495\textwidth}{(a)}
\fig{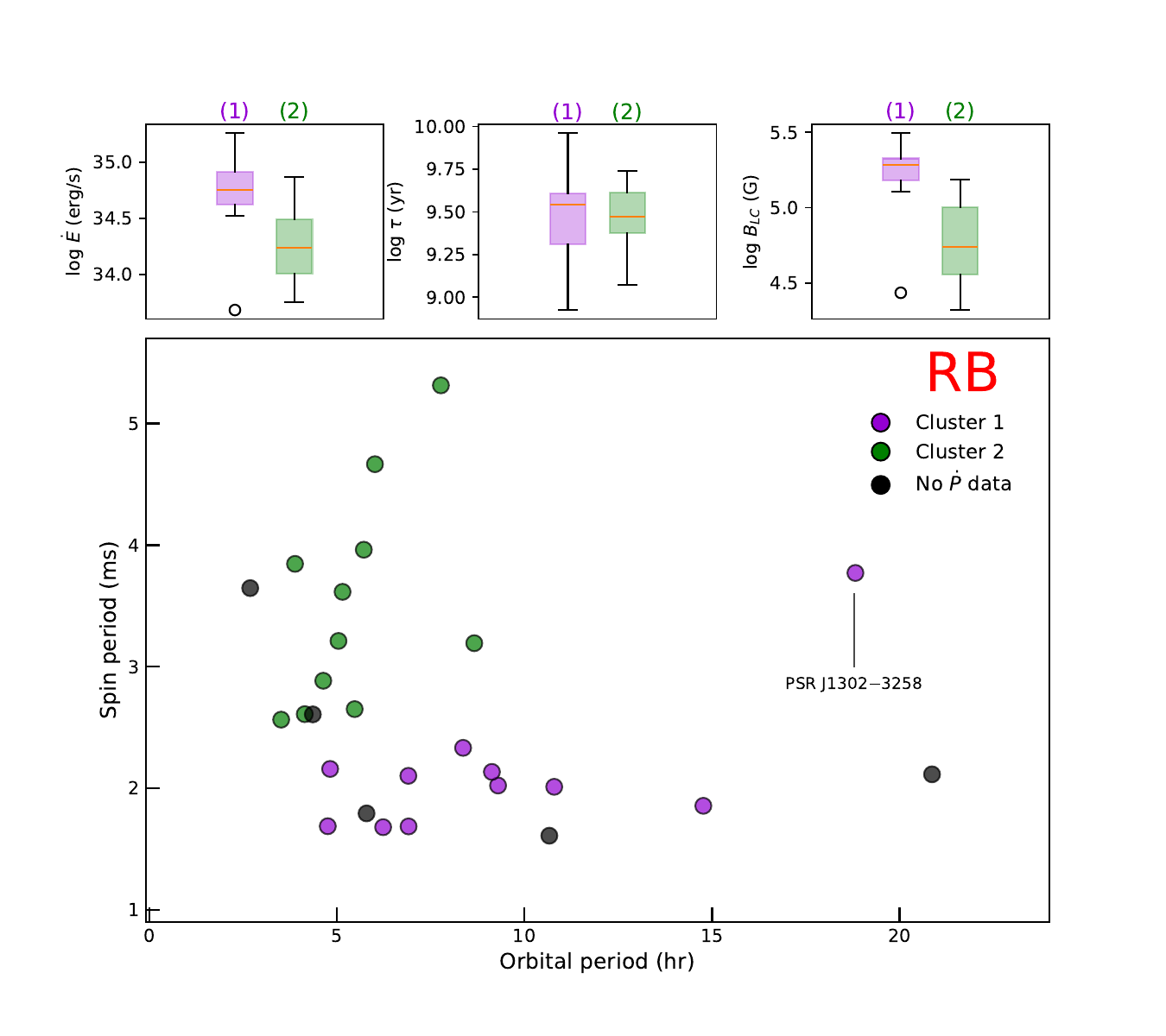}{0.495\textwidth}{(b)}}
\caption{Spider spin periods as a function of orbital period (panel (a): RBs, panel (b): BWs). Data points are color coded by K-means clusters (black indicates no available estimate for $\dot{P}$): Cluster 1 (violet data points) features fast spins and a wider range in orbital periods, while Cluster 2 (green data points) comprises systems with short orbital periods and a broader range of spin periods. Upper panels display box plots of derived parameters for the two clusters: spin-down luminosity (left), characteristic age (middle), and magnetic field strength at the light cylinder (right). 
\label{fig:spin_orbit}}
\end{figure*}

Spiders appear to cluster along two axes in the spin period versus orbital period plane (Figure~\ref{fig:spin_orbit}). The majority of sources fall into one of two groups: either fast-spinning pulsars ($<$3~ms) with a wide range of orbital periods, or systems with short orbital periods ($\sim$3~hr for BWs, $\sim$5~hr for RBs) and a broader range of spin periods (2--5~ms). 

To quantify this, we performed a K-means clustering on a spider-type separated dataset including parameters $P$, $\dot{P}$, $P_{\rm b}$, and the projected semi-major axis of the orbit ($A_1$) acquired from ATNF. From this analysis, we exclude HMs and two long spin period sources: PSR~J1932$+$2121 and PSR~J2129$-$0429, which are mildly recycled pulsars. The results for two clusters are shown in Figure~\ref{fig:spin_orbit} for both RBs and BWs, possibly indicating differing source populations. We also derived distributions for the spin-down luminosity, characteristic age, and the magnetic strength at the light cylinder for both clusters (top panels in Figure~\ref{fig:spin_orbit}). Here, we defined the magnetic field strength at the light cylinder as $B_{\rm LC} = B(R/R_{\rm LC})^3$, where $B$ is the surface magnetic field strength defined in Section~\ref{sec:spins}, $R=10$ km, and $R_{\rm LC} = cP/2\pi$.

These clusters also differ significantly in spin-down luminosity ($\sigma\sim2.6-3.1$),\footnote{Medians $\bar{\dot{E}}_{\rm RB, 1}=10^{34.8}$ erg s$^{-1}$ and $\bar{\dot{E}}_{\rm RB, 2}=10^{34.2}$ erg s$^{-1}$ (Mann-Whitney $U=101$, $n_1=11$, $n_2=11$, $p<0.009$, two-tailed) for RBs; $\bar{\dot{E}}_{\rm BW, 1}=10^{34.8}$ erg s$^{-1}$ and $\bar{\dot{E}}_{\rm BW, 2}=10^{34.2}$ erg s$^{-1}$ (Mann-Whitney $U=265$, $n_1=13$, $n_2=25$, $p<0.002$, two-tailed) for BWs.} and magnetic field strength at the light cylinder ($\sigma\sim3.1-3.5$),\footnote{Medians $\bar{B}_{\rm LC, RB, 1}=10^{5.3}$ G and $\bar{B}_{\rm LC, RB, 2}=10^{4.7}$ G (Mann-Whitney $U=109$, $n_1=11$, $n_2=11$, $p<0.002$, two-tailed) for RBs; $\bar{B}_{\rm LC, BW, 1}=10^{5.2}$ G and $\bar{B}_{\rm LC, BW, 2}=10^{4.8}$ G (Mann-Whitney $U=277$, $n_1=13$, $n_2=25$, $p<0.0005$, two-tailed) for BWs.} but not in characteristic age ($\sigma\sim0.3-1.5$),\footnote{Medians $\bar{\tau}_{\rm RB, 1}=10^{9.5}$ yr and $\bar{\tau}_{\rm RB, 2}=10^{9.5}$ yr (Mann-Whitney $U=56$, $n_1=11$, $n_2=11$, $p<0.8$, two-tailed) for RBs; $\bar{\tau}_{\rm BW, 1}=10^{9.5}$ yr and $\bar{\tau}_{\rm BW, 2}=10^{9.7}$ yr (Mann-Whitney $U=113$, $n_1=13$, $n_2=25$, $p<0.13$, two-tailed) for BWs.} as shown in the upper panels of Figure~\ref{fig:spin_orbit}. Overall, compared to Cluster 2, Cluster 1 exhibits higher spin-down luminosities and stronger magnetic fields at the light cylinder. These trends are consistent across both BWs (Figure~\ref{fig:spin_orbit}, panel (a)) and RBs (Figure~\ref{fig:spin_orbit}, panel (b)). 

According to binary evolution models of spider systems, slower spin periods may result from a lower accretion efficiency during the recycling phase or from a higher initial neutron star mass \citep[e.g.,][and references therein]{tauris12,chen13,kar24,misra25a}. These long-$P$ systems would also exhibit lower spin-down luminosities ($\dot{E} \propto P^{-3}$), leading to reduced companion irradiation. Weaker irradiation inhibits mass loss from the binary, thereby limiting orbital widening. As a result, Cluster 2 spiders would maintain a narrow range of orbital periods while showing a wider distribution of spin periods, shaped by variations in initial neutron star mass or accretion efficiency. In contrast, spiders in Cluster 1 likely underwent more efficient accretion during recycling phase, reaching spin periods of $\sim$2~ms. These systems display higher spin-down luminosities and likely stronger companion irradiation, which lead to wider orbits and a broader range of orbital periods depending on irradiation efficiency.

We note potential outliers for the clustering. Among the RBs, PSR~J1302$-$3258 has $P_{\rm b}=18.8$~hr, $P=3.77$~ms, and $M_{\rm c,min} = 0.15 M_{\odot}$, standing out in the spin-orbit plane (indicated in Figure~\ref{fig:spin_orbit}, panel (b)) and presenting very different derived parameters from Cluster 1 sources ($\dot{E} = 5 \times 10^{33}$ erg s$^{-1}$ and $B_{\rm LC} = 2.7 \times 10^4$ G) and shows eclipses at low radio frequencies, fulfilling our RB definition (Section~\ref{sec:definition}). However, the eclipses last for only $\sim$10\% of the orbit, and the system lacks an optical counterpart despite its apparent proximity ($d \approx 1.4$ kpc from DM and YMW16, Section~\ref{sec:distances}), raising doubts about its RB identification \citep{bbc+24}.

A few BWs can be found at intermediate spins (3--4~ms) and orbital periods (6--10~hr) offset from most sources. These are PSR~J2234$+$0944, PSR~J0610$-$2100, PSR~J1630$+$3550, PSR~J2214$+$3000, and PSR~J0541$+$2959g. All exhibit low spin-down luminosities ($\dot{E} <2 \times 10^{34}$ erg s$^{-1}$; no estimate exists for PSR~J0541$+$2959g due to an unmeasured $\dot{P}$), in line with Cluster 2 sources. Therefore, their higher orbital periods are unlikely to result from irradiation effects. Notably, three of these sources---PSRs J0610$-$2100, J2214$+$3000, and J2234$+$0944---are the only spiders included in pulsar timing arrays (NANOGrav: \citealt{arzoumanian18}; EPTA: \citealt{desvignes16,baknielsen20}), demonstrating remarkably stable timing over a decade \citep{vbc+22}. This is atypical, as most spiders show timing irregularities on timescales of months to a few years \citep[e.g.,][]{aft94,pc15,deneva16,svf+16,cnv+21}, likely caused by changes in the companion star's gravitational quadrupole moment \citep{applegate87,applegate94} or tidal forces from asynchronous rotation \citep{vanstaden16}. While not unique among BWs, these outliers show no or only intermittent, low-frequency radio eclipses \citep{bgc+13,lod+23,kumari25,why+24}, suggesting either an absence or minimal presence of ionized material in the system, or that such material does not intersect the line of sight. One possibility is that the companions are semidegenerate or degenerate stars, which are more resistant to ablate.

\section{Conclusions} \label{sec:conclusions}

The study of compact binary MSPs is experiencing a remarkable surge, driven by an increasing number of discoveries and their importance for fundamental astrophysics. This rapidly growing class of pulsars, known as spiders, consists of neutron stars in tight binary orbits that gradually strip material from their low-mass companions. These systems are among the best laboratories for studying pulsar evolution and are key to identifying the most massive neutron stars, which provide crucial constraints on the equation of state of ultradense matter.

The number of known spider systems and candidates in the Galactic field has now surpassed 100, marking the transition into an era of population-level statistical studies. In addition to the two main subclasses, several peculiar systems and subclasses have emerged, including mildly recycled pulsars with slow spin periods, transitional systems that alternate between accretion-powered and rotation-powered states (tMSPs), and those with unusual companion stars, such as extremely low-mass companions (tidarren) or giant companions (HM).

In this paper, we compiled and summarized multiwavelength data and radio timing parameters for the current population of spider systems. The paper is accompanied by a public online database, \footnote{\url{https://astro.phys.ntnu.no/SpiderCAT}} which we aim to update continuously as new sources are discovered or existing entries are refined.\footnote{We welcome community feedback, see \url{https://astro.phys.ntnu.no/SpiderCAT/acknowledgements}.} Tables 2--5 are also available in machine-readable form in Vizier.\footnote{\url{https://cdsarc.cds.unistra.fr/viz-bin/cat/J/ApJ/994/8}} The catalog highlights the most relevant parameters of each system, as presented here.

Utilizing the catalog, we found the following. 
\begin{enumerate}
    \item Confirmed spiders (excluding mildly recycled systems) have spin periods between 1.4 and 5.4 ms and spin period derivatives ranging from $10^{-21}$ to $7.5 \times 10^{-20}$ s s$^{-1}$.
    \item Excluding tidarren and HM systems, the orbital periods range from approximately 2 to 26 hr. Only four RBs (12.5\%) have orbital periods shorter than 4 hr, compared to 27 BWs (55\%).
    \item Parameters observed exclusively in RB systems include minimum companion masses greater than $M_{\rm c, min} > 0.06,M_{\odot}$, X-ray luminosities exceeding $L_X > 1.3 \times 10^{32}$ erg s$^{-1}$, and $\gamma$-ray to X-ray flux ratios below $\log (F_{\gamma}/F_X) < 2$.
    \item In a color-magnitude diagram, most spiders lie between the main sequence and the WD branch. In the Pan-STARRS filters, this region is approximately bounded from above by $M_g = 4.5 \times (g - r) + 4.5$ and from below by $M_g = 2.7 \times (g - r) + 10.3$, providing a useful criterion for identifying spider candidates in optical surveys. Notably, RBs exhibit absolute $g$-band magnitudes roughly 2 mag brighter than BWs at similar colors.
    \item 
    The spider distribution is peaked toward the Galactic plane (Figure~\ref{fig:skymap}). We estimate a Galactic scale height scale (Figure~\ref{fig:galactic_height}) of $z_{e}=0.73\pm0.15$ kpc for spiders.
\end{enumerate}  

The future of spider searches remains bright. Many spiders are likely hidden in the Galactic plane---a region historically underexplored due to source confusion (with the notable recent exception of the FAST-GPPS survey). Given the relatively high Galactic scale height of spiders, many may also remain undetected at higher latitudes.
Additionally, a large number of unassociated Fermi sources remain to be classified \citep[$>$ 2000;][]{aab+22, mayer24}, many of which may host a spider pulsar.
Therefore, we anticipate the number of known Galactic-field spiders to continue increasing. 

%% Please use the acknowledgment and contribution environments. This will 
%% be anonomyized when the "anonymous" style option is used. 
\begin{acknowledgments}

% Personal thanks
We thank Iacob Nedreaas and Bogdan Voaidas for software contributions to the catalog's search engine and web server. We are also grateful to Devina Misra for discussions on the binary evolutionary models of spiders, and to the anonymous referee for comments that improved the paper. In addition, we thank Cl\'ement Vidal for pointing out an error in the Gaia DR3 ID numbers.

% LOVE-NEST
This project has received funding from the European Research Council (ERC) under the European Union’s Horizon 2020 research and innovation programme (grant agreement No. 101002352, PI: M. Linares).

% 2MASS
This publication makes use of data products from the Two Micron All Sky Survey, which is a joint project of the University of Massachusetts and the Infrared Processing and Analysis Center/California Institute of Technology, funded by the National Aeronautics and Space Administration and the National Science Foundation.

% Pan-STARRS
The Pan-STARRS1 Surveys (PS1) and the PS1 public science archive have been made possible through contributions by the Institute for Astronomy, the University of Hawaii, the Pan-STARRS Project Office, the Max-Planck Society and its participating institutes, the Max Planck Institute for Astronomy, Heidelberg and the Max Planck Institute for Extraterrestrial Physics, Garching, The Johns Hopkins University, Durham University, the University of Edinburgh, the Queen's University Belfast, the Harvard-Smithsonian Center for Astrophysics, the Las Cumbres Observatory Global Telescope Network Incorporated, the National Central University of Taiwan, the Space Telescope Science Institute, the National Aeronautics and Space Administration under Grant No. NNX08AR22G issued through the Planetary Science Division of the NASA Science Mission Directorate, the National Science Foundation Grant No. AST-1238877, the University of Maryland, Eotvos Lorand University (ELTE), the Los Alamos National Laboratory, and the Gordon and Betty Moore Foundation.

% SkyMapper
The national facility capability for SkyMapper has been funded through ARC LIEF grant LE130100104 from the Australian Research Council, awarded to the University of Sydney, the Australian National University, Swinburne University of Technology, the University of Queensland, the University of Western Australia, the University of Melbourne, Curtin University of Technology, Monash University and the Australian Astronomical Observatory. SkyMapper is owned and operated by The Australian National University's Research School of Astronomy and Astrophysics. The survey data were processed and provided by the SkyMapper Team at ANU. The SkyMapper node of the All-Sky Virtual Observatory (ASVO) is hosted at the National Computational Infrastructure (NCI). Development and support of the SkyMapper node of the ASVO has been funded in part by Astronomy Australia Limited (AAL) and the Australian Government through the Commonwealth's Education Investment Fund (EIF) and National Collaborative Research Infrastructure Strategy (NCRIS), particularly the National eResearch Collaboration Tools and Resources (NeCTAR) and the Australian National Data Service Projects (ANDS).

% eROSITA
 This work is based on data from eROSITA, the soft X-ray instrument aboard SRG, a joint Russian-German science mission supported by the Russian Space Agency (Roskosmos), in the interests of the Russian Academy of Sciences represented by its Space Research Institute (IKI), and the Deutsches Zentrum für Luft- und Raumfahrt (DLR). The SRG spacecraft was built by Lavochkin Association (NPOL) and its subcontractors, and is operated  by NPOL with support from the Max Planck Institute for Extraterrestrial Physics (MPE). The development and construction of the eROSITA X-ray instrument was led by MPE, with contributions from the Dr. Karl Remeis Observatory Bamberg \& ECAP (FAU Erlangen-Nuernberg), the University of Hamburg Observatory, the Leibniz Institute for Astrophysics Potsdam (AIP), and the Institute for Astronomy and Astrophysics of the University of Tübingen, with the support of DLR and the Max Planck Society. The Argelander Institute for Astronomy of the University of Bonn and the Ludwig Maximilians Universität Munich also participated in the science preparation for eROSITA. 

 % XMM
 This research has made use of data obtained from the 4XMM XMM-Newton serendipitous source catalogue compiled by the XMM-Newton Survey Science Centre consortium.
 
\end{acknowledgments}

\begin{contribution}
%%This section gives authors the space to recognize author contributions. The text inside this environment is NOT counted towards the total word quanta. At a minimum, manuscripts are expected to include this text:

K.K. led the development of this work, including writing and preparing the original manuscript draft, designing and implementing the software tools, curating and analyzing the data set, generating the visualizations, and conducting the formal analysis. K.K. was also responsible for the submission process and the overall coordination of manuscript preparation.

M.L. conceived the original idea for the project and contributed to the spider definition, data curation and software development. M.L. reviewed, edited, and contributed to the manuscript throughout its preparation, secured the funding necessary to support the research and manages the project’s GitHub repository.

%% But authors are expected to provide more specific details, e.g. 
%%
%%SC was responsible for writing and submitting the manuscript.
%%WWM came up with the initial research concept and edited the manuscript.
%%OTS obtained the funding and edited the manuscript.
%%EBF provided the formal analysis and validation. He also edited the manuscript.
%%GEH Supervised the undergraduates, wrote the software and administers the project github and Zenodo repositories.
%%
%% Authors can use the Contributor Role Taxonomy (CRediT) at
%% for ideas on how write a good statement tailored to their needs.

\end{contribution}

\facilities{Gaia, CTIO:2MASS, FLWO:2MASS, PS1, Skymapper, Swift(XRT), CXO, eROSITA, XMM, Fermi}

\software{
          VizieR \citep{ochsenbein00},
          HEASoft/HI4PI \citep{HI4PI},  
          PyGEDM \citep{price21},
          Bayestar \citep{green19}, 
          astropy \citep{astropy1,astropy2},
          psrqpy \citep{psrqpy},
          astroquery \citep{astroquery}
          dustmaps \citep{dustmaps}
          }


%\appendix

%\section{Appendix A}
%\section{Appendix B}

%% For this sample we use BibTeX plus aasjournalv7.bst to generate the

\bibliography{biblio_spidercat}{}
\bibliographystyle{aasjournalv7}



\end{document}